\documentclass[traditabstract]{aa}
\usepackage{natbib}
\bibpunct{(}{)}{;}{a}{}{,} % to follow the A&A style
\usepackage{subfig}
\usepackage{amsmath}
\usepackage{enumerate}
\usepackage{aas_macros}
\usepackage{graphicx}
\usepackage{txfonts}

\usepackage{amssymb}

\SetSymbolFont{symbols}{bold}{OMS}{cmsy}{b}{n}
\DeclareSymbolFont{bmisymbols}{OML}{cmm}{b}{it}
\DeclareMathSymbol{\balpha}{0}{bmisymbols}{"0B}
\DeclareMathSymbol{\bbeta}{0}{bmisymbols}{"0C}
\DeclareMathSymbol{\bgamma}{0}{bmisymbols}{"0D}
\DeclareMathSymbol{\bdelta}{0}{bmisymbols}{"0E}
\DeclareMathSymbol{\bepsilon}{0}{bmisymbols}{"0F}
\DeclareMathSymbol{\bzeta}{0}{bmisymbols}{"10}
\DeclareMathSymbol{\boldeta}{0}{bmisymbols}{"11}
\DeclareMathSymbol{\btheta}{0}{bmisymbols}{"12}
\DeclareMathSymbol{\biota}{0}{bmisymbols}{"13}
\DeclareMathSymbol{\bkappa}{0}{bmisymbols}{"14}
\DeclareMathSymbol{\blambda}{0}{bmisymbols}{"15}
\DeclareMathSymbol{\bmu}{0}{bmisymbols}{"16}
\DeclareMathSymbol{\bnu}{0}{bmisymbols}{"17}
\DeclareMathSymbol{\bxi}{0}{bmisymbols}{"18}
\DeclareMathSymbol{\bpi}{0}{bmisymbols}{"19}
\DeclareMathSymbol{\brho}{0}{bmisymbols}{"1A}
\DeclareMathSymbol{\bsigma}{0}{bmisymbols}{"1B}
\DeclareMathSymbol{\btau}{0}{bmisymbols}{"1C}
\DeclareMathSymbol{\bupsilon}{0}{bmisymbols}{"1D}
\DeclareMathSymbol{\bphi}{0}{bmisymbols}{"1E}
\DeclareMathSymbol{\bchi}{0}{bmisymbols}{"1F}
\DeclareMathSymbol{\bpsi}{0}{bmisymbols}{"20}
\DeclareMathSymbol{\bomega}{0}{bmisymbols}{"21}
\DeclareMathSymbol{\bvarepsilon}{0}{bmisymbols}{"22}
\DeclareMathSymbol{\bvartheta}{0}{bmisymbols}{"23}
\DeclareMathSymbol{\bvarpi}{0}{bmisymbols}{"24}
\DeclareMathSymbol{\bvarrho}{0}{bmisymbols}{"25}
\DeclareMathSymbol{\bvarsigma}{0}{bmisymbols}{"26}
\DeclareMathSymbol{\bvarphi}{0}{bmisymbols}{"27}

\newcommand{\rmn}{\mathrm}
\newcommand{\dd}{\mathrm{d}}

\newcommand{\vel}{\upsilon}
\newcommand{\vvel}{\bupsilon}
\newcommand{\vst}{\vel_\rmn{st}}
\newcommand{\vvst}{\vvel_\rmn{st}}

\newcommand{\p}{\rmn{p}}
\newcommand{\e}{\rmn{e}}
\newcommand{\eps}{\varepsilon}

\title{Cosmic ray transport in galaxy clusters:\\
implications for radio halos, gamma-ray signatures,\\
and cool core heating}
\titlerunning{Cosmic ray transport in galaxy clusters}
\author{Torsten En{\ss}lin\inst{1} \and Christoph Pfrommer\inst{2} \and Francesco Miniati\inst{3} \and Kandaswamy Subramanian\inst{4}}
\authorrunning{En{\ss}lin et al.}
\institute{Max-Planck-Institut f\"ur Astrophysik, Karl-Schwarzschild-Str. 1, D-85741 Garching bei M\"unchen, Germany
\and 
Canadian Institute for Theoretical Astrophysics,
60 St. George Street
Toronto, Ontario, M5S 3H8, Canada
\and 
ETH Zurich Institute of Astronomy, Physics Department, HIT J 12.2. Wolfgang-Pauli-Strasse 27. CH-8093 Zurich, Switzerland
\and 
Inter-University Centre for Astronomy \& Astrophysics, Post Bag 4, Ganeshkhind, Pune 411 007, India
}
\date{Submitted 27.8.2010}
\begin{document}
%{\onecolumn
%\maketitle
%\begin{abstract}
\abstract{ We investigate the interplay of cosmic ray (CR) propagation
  and advection in galaxy clusters. Propagation in form of CR
  diffusion and streaming tends to drive the CR radial profiles
  towards being flat, with equal CR number density
  everywhere. Advection of CR by the turbulent gas motions tends to
  produce centrally enhanced profiles.  We assume that the CR streaming 
  velocity is of the order of the sound velocity. This is motivated by 
  plasma physical arguments. The CR streaming is then usually larger than 
  typical advection velocities and becomes comparable or lower than this only for periods 
  with trans- and super-sonic cluster turbulence. As a consequence
  a bimodality of the CR spatial distribution results. Strongly turbulent, merging
  clusters should have a more centrally concentrated CR energy density
  profile with respect to relaxed ones with very subsonic turbulence.
  This translates into a bimodality of the expected diffuse radio and
  gamma-ray emission of clusters, since more centrally concentrated CR
  will find higher target densities for hadronic CR proton
  interactions, higher plasma wave energy densities for CR electron
  and proton re-acceleration, and stronger magnetic fields.  Thus, the
  observed bimodality of cluster radio halos appears to be a natural
  consequence of the interplay of CR transport processes, independent
  of the model of radio halo formation, be it hadronic interactions of
  CR protons or re-acceleration of low-energy CR
  electrons. 
  Energy dependence of the CR propagation should lead to spectral steepening of dying radio halos.
  Furthermore, we show that the interplay of CR diffusion
  with advection implies first order CR re-acceleration in the
  pressure-stratified atmospheres of galaxy clusters.  Finally, we
  argue that CR streaming could be important in turbulent cool cores
  of galaxy clusters since it heats preferentially the central gas
  with highest cooling rate.  }
%\end{abstract}
%}

\maketitle

%\glossary{name={entry name}, description={entry description}}

\section{Introduction}
\label{sec:intro}
\subsection{Motivation}

Relativistic particle populations, cosmic rays (CR),  are expected to permeate the intra-cluster medium (ICM). Cosmic ray electrons (CRe) are directly visible in many galaxy clusters via their radio synchrotron emission, forming the so-called cluster radio halos \citep[e.g.][]{2004rcfg.procE..25K, 2004NewAR..48.1137F}. Several CRe injection sites can also be identified via the same synchrotron radiation mechanism: winds and gas stripping from cluster galaxies, active galactic nuclei, and shock waves from structure formation. All these should also be injection sites for CR protons (CRp) and heavier relativistic nuclei. Due to their higher masses with respect to the electrons, protons and nuclei are accelerated more efficiently. In our own Galaxy, the ratio of the spectral energy flux of CRp to CRe between 1\ldots 10 GeV is about one 
hundred~\citep{2002cra..book.....S}. Similar ratios are also expected at least for the injection from galaxies and structure formation shock waves for the same kinematic reasons. 

Cluster CRp should have accumulated over cosmic timescales since the
bulk of them is unable to leave through the persistent infall of
matter onto the cluster and due to the long CRp's radiative lifetimes
in the ICM of the order of an Hubble time \citep{1996SSRv...75..279V,
  1997ApJ...477..560E, 1997ApJ...487..529B}. CRe suffer much more
severe energy losses via synchrotron and inverse Compton emission at
GeV energies, and Bremsstrahlung and Coulomb losses below 100 MeV. CRe
with an energy of $\sim 10~$GeV emit GHz synchrotron waves in
$\mu$G-strength magnetic fields. Since the associated inverse Compton
and synchrotron cooling time is $\tau_\rmn{IC,syn}\sim 2\times
10^8$~yr, these CRe must have been recently injected or
re-accelerated, whereas any CRp can be as old as its cluster. Since
CRp should be abundant and are able to inject relativistic electrons
via the production of charged pions in hadronic interactions with gas
nuclei, they could be the origin of the observed radio
halos\footnote{\label{note:hadronic}\citet{1980ApJ...239L..93D,1982AJ.....87.1266V, 1999APh....12..169B, 2000A&A...362..151D, 2001ApJ...559...59M, 2003MNRAS.342.1009M, 2004A&A...413...17P, 2004MNRAS.352...76P, 2008MNRAS.385.1211P, 2008MNRAS.385.1242P,  2009JCAP...09..024K, 2010MNRAS.401...47D, 2010arXiv1003.0336D, 2010arXiv1003.1133K, 2010arXiv1011.0729K}}. 
Alternatively, a low energy (100 MeV) CRe ICM population might be sufficiently long lived if it is maintained by re-acceleration by plasma waves against cooling processes. During the phases of high ICM turbulence after cluster merger the re-acceleration might be so efficient that CRe are accelerated into the radio observable energy range of $\sim 10$~GeV.\footnote{\label{note:reacc}\citet{1987A&A...182...21S, 1993ApJ...406..399G, 2004MNRAS.350.1174B, 2005MNRAS.363.1173B, 2007MNRAS.378..245B, 2010arXiv1008.0184B, 2009A&A...507..661B}}

For both particle populations, CRp\footnote{Actually, there should also be a substantial population of relativistic alpha-particle. These can, however, regarded to be equivalent to an ensemble of four CRp, traveling together due to the relatively weak nuclear binding forces between them (and keeping two of them as neutrons). See \citet{2007A&A...473...41E} for an extended discussion of this.} and CRe, which we address commonly as CR, the question is what spatial and spectral shape they have acquired? 
This will largely determine which radiative signatures and which dynamical influences we can expect from their presence. 

Without continuous injection of CRe, their spectra should be highly curved, due to the severe energy losses relativistic electrons suffer from, both at the high and low end of their spectra. If not replenished, or sufficiently re-accelerated, the CRe population should disappear in the cluster center after a time of about a Gyr. Thus, the CRe visible in the radio halos should be spatially close to their injection or last re-acceleration site.

At energies above a few GeV, the injected CRp power law spectra are
unaffected by hadronic and adiabatic losses, only the normalisation of
the spectra evolves
\citep[e.g.,][]{2001CoPhC.141...17M,2007A&A...473...41E,
  2010arXiv1001.5023P}. The spatial distribution, however, may be
strongly affected by transport processes. 
Also local spectral modifications can be expected if the macroscopic transport mechanism depends on the CR energy.
Although the CRp travel
close to the speed of light, the ICM magnetic fields of
$\mathcal{O}(\mu$G$)$ force them on helical orbits along the field
lines, and resonant scattering events with plasma waves will try to isotropise the CR momentum distribution rapidly. Thus CRp are tied to the fields and their transport is strongly controlled by the magnetic and turbulent properties of the ICM.

In addition to the radiative signatures of CRs in the radio and gamma
ray bands, CRs can store significant amount of energy, which they
preferentially release in the centers of clusters due to the higher target density there and because of the adiabatic losses they suffer when they propagate to the cluster outskirts. Thus, CRs were proposed to help to stabilise cluster cool cores against a cooling instability\footnote{\label{note:coolcoreheat}Heating via Coulomb losses of CRs from a central AGN are discussed in \citet{2004A&A...413..441C} and \citet{2008A&A...484...51C}. The additional heating via streaming of the same CRs is considered in \citet{2008MNRAS.384..251G}.}. The spatial distribution of the CR population, as shaped by transport processes, as well as the energy absorbed and released during advective and streaming transport are of direct importance for this.
Three transport processes are relevant here:
\begin{itemize}
 \item \textbf{Advection:} The magnetic field lines are largely
frozen into the thermal plasma of the ICM and are dragged with any gas flow. The enclosed CR are advected with this flow and suffer energy losses or gains from any adiabatic expansion or compression of the flow, respectively.
\item \textbf{Diffusion:} A CR may travel several gyro-radii along a field line before it is resonantly scattered by plasma waves of the medium. 
The resulting random walk along the field line leads to a considerable diffusion parallel to the field lines. 
Since the gyro center of the particles are displaced in the plasma-wave interaction, a small perpendicular diffusion results. 
This leads to a larger macroscopic displacement from the original magnetic field line when the particles follow the diverging path of this initially neighboring line. 
The CR diffusion coefficient generally increases with the particle energy.
\item \textbf{Streaming:}  In the presence of a sufficiently large
  gradient in the spatial CR distribution along a field line, an
  anisotropic momentum distribution function builds up, since more
  particle arrive at such a location from one side than from the
  other. This leads to a net CR flux towards the CR depleted
  region. The streaming of the CR with respect to the thermal ICM
  excites plasma waves, on which the particles scatter. This limits
  the streaming velocity of CRs to be of the order of the Alfv\'enic
  or the sound speed, depending on the plasma properties.
\end{itemize}
Advection is a passive form of transport, which is actually included in many of the numerical simulation schemes for CRp in the large-scale-structure\footnote{\citet{2001CoPhC.141...17M,2002MNRAS.337..199M,2007JCoPh.227..776M,2007ApJ...667L...1M,2006MNRAS.367..113P, 2007A&A...473...41E, 2008A&A...481...33J}}. Diffusion and streaming are active propagation processes, relying on the own speed of the CR. Inclusion of CR diffusion exists for some MHD codes\footnote{\citet{2002CoPhC.147..471B, 2003and..book..269B, 2003A&A...412..331H, 2006MNRAS.373..643S, 2010EAS....42..275H, 2010EAS....42..281H}}, which are, however, not suited for large-scale-structure simulations. To our knowledge, CR streaming is not implemented in any of the simulation codes used in astrophysics. Thus, the effects we discuss in this paper concern physics, which is not captured in current numerical simulations of galaxy clusters.

\subsection{Goal and outline}

With this paper, we want to discuss the possible effects CR
propagation has on the spatial and spectral distribution of CR in
galaxy clusters and their radiative signatures in gamma-rays and radio
frequencies. At this stage of the research, without the necessary cosmological simulation tools including simultaneously CR propagation and magnetic field evolution at hand, it is not possible to make definite predictions, due to the complexity of the interaction of turbulent ICM gas with CR and magnetic fields. Our goal here is to outline plausible scenarios, which highlight the potential importance of CR propagation, and thereby motivates further research and hopefully lead to the development of the necessary simulation tools. 

A strong motivation for such developments should be the fact that CR propagation can explain the observed bimodality of cluster radio halos within the hadronic radio halo model. 
Since all necessary ingredients of this model, as CRp shock acceleration, a long CRp cooling times, the processes of hadronic $e^\pm$ production and $e^\pm$ synchrotron losses, as well as the required field strength of a few $\mu$G, are known to be given in typical ICM environments, this model is very natural, nearly free of assumptions and therefore attractive. 
The main counter-argument, that not every cluster exhibits a radio halo, is alleviated if CR propagation is operative in the ICM as argued here. 
However, also radio halos in the CRe re-acceleration model should
strongly benefit from propagation effects. The low energy CRe
population, which is re-accelerated to explain radio halos in these
models, is probably thermalised in the cluster centers due to Coulomb energy losses during 
quiet phases of the cluster, when re-acceleration is weak. However, CRe
can survive for Gyrs in the outskirts of clusters during such phases, and being dragged into the core during later turbulent merger phases. Then, the required turbulent re-acceleration is also present, in order to power a radio halo in the re-acceleration model.
Also for this model, the observed radio bimodality of clusters should partly result from the interplay of the CRe transport mechanisms.

The outline of this work is the following.  In
Sect. \ref{sec:radiohalos}, we first introduce radio halos, their
connection to the dynamical state of clusters, and the strength and
weaknesses of the two main theoretical scenarios used to explain
them. Then we discuss in Sect. \ref{sec:CRtransp} typical conditions
of the turbulent ICM and their implications for CR transport processes
and the implied adiabatic energy gains and losses. The consequences
for non-thermal cluster emission in the gamma-ray and radio spectral
bands as well as the potential heating of cool cores are investigated in Sect. \ref{sec:implication}. Finally we conclude in Sect. \ref{sec:discuss}.

\section{Radio halos}
\label{sec:radiohalos}
\subsection{Observational properties}
\label{sec:haloprop}

\begin{figure*}
 \centering
 \includegraphics[bb=135 253 478 534, width=0.49\textwidth]{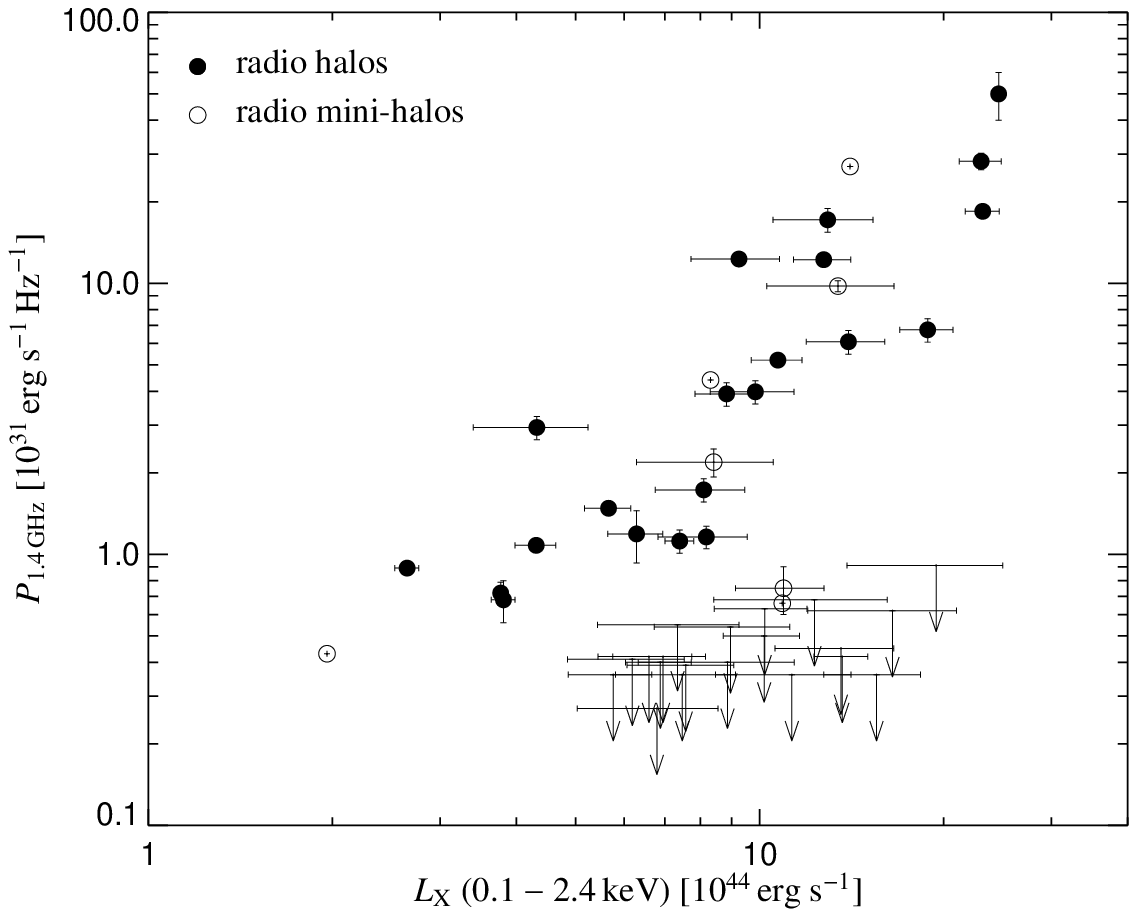}
 % halo.eps: 0x0 pixel, 300dpi, 0.00x0.00 cm, bb=135 253 478 534
\includegraphics[bb=135 253 478 534, width=0.49\textwidth]{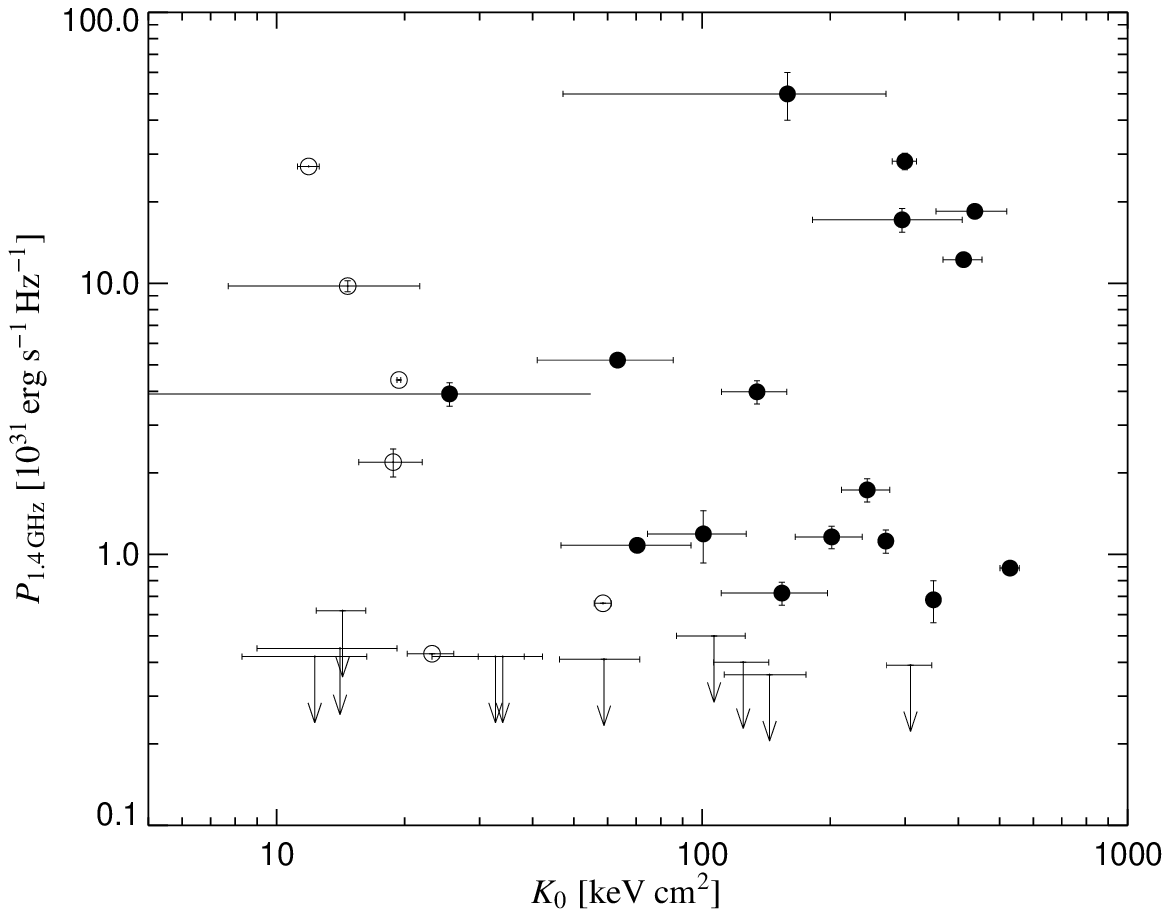}
% halo_entropy.eps: 0x0 pixel, 300dpi, 0.00x0.00 cm, bb=135 253 478 534
 \caption{Correlation of radio halo luminosities with cluster properties of clusters in Tab. \ref{tab:clusterstat}.
\textbf{Left:} Radio halo luminosity vs X-ray luminosity. 
\textbf{Right:} Radio halo luminosity vs central entropy indicator $K_0$ for the subsample of clusters for which high resolution Chandra data are available.
}
 \label{fig:radiohalostat}
\end{figure*}

\begin{table*}
%\begin{singlespace}  %%  only to match referee format
%\vspace{-2em} %%  only to match referee format
 \begin{center}
% use packages: array
\begin{tabular}{|l|llll|l|}
% radio:
% base sample from Brunetti, Cassano, Dolag, Setti 2009,
% 4 mini-halos from Gitti, Brunetti, Feretti, Setti (2004)
%
% X-ray:
% LX of last 4 mini-halos from Reiprich & Boehringer 2002: L_X (0.1-2.4 keV) [h50^-2 * (0.5/0.7)^2 1e44 erg/s]
% L_X of A2626 from Stott et al 2008, MNRAS, 384, 1502
%
% central entropies K0 (using the extrapolation method on K0 + K100 * (r/100 kpc)^alpha to obtain K0)
% base sample from Cavagnolo, Donahue, Voit, Sun, 2009, ApJS, 182, 12
% K0 of Coma @12 kpc, from Rafferty, McNamara, Nulsen, 2008, ApJ, 687, 899
%
% columns:
% 1 ID
% 2 redshift z
% 3 L_X (0.1-2.4 keV) [1e44 erg/s] 
% 4 Delta L_X (0.1-2.4 keV) [1e44 erg/s] 
% 5 K0 (ext)           [keV cm^2]
% 6 Delta K0 (1 sigma) [keV cm^2]
% 7 P_{1.4 GHz}       [10^24 W/Hz]
% 8 Delta P_{1.4 GHz} [10^24 W/Hz]
%
\hline
 & redshift & $L_\rmn{X}$ & ${K_0}$& ${P_\rmn{1.4 GHz}}$ & Reference\\
 & & ${10^{44}\,\rmn{erg~s}^{-1}}$ & ${\rmn{keV~cm}^2}$ & ${10^{31} \rmn{erg~s}^{-1}\,\rmn{Hz}^{-1}}$ & \\
\hline
giant radio halos & & & &  &  \\
\hline
1E50657-558    & 0.2994  & 23.03 $\pm$ 1.81  & 299.4 $\pm$ 19.6   & 28.21 $\pm$   1.97 & $4$, $24$\\  % all giant halos
A209           & 0.2060  & 6.29  $\pm$ 0.65  & 100.7  $\pm$  26.3 & 1.19  $\pm$
0.26  & $4$,$28$\\
A520 	       &  0.2010 & 8.83  $\pm$ 1.99  & 325.5  $\pm$  29.2 & 3.91  $\pm$  0.39  & $12$, $2$\\
A521           & 0.2475  & 8.18  $\pm$ 1.36  & 201.6   $\pm$ 36.1 &  1.16  $\pm$  0.11 & $4$, $9$\\
A545 	       & 0.1530  & 5.66  $\pm$ 0.49  &    --              & 1.48   $\pm$ 0.06  & $4$, $2$\\
A665           & 0.1816  & 9.84  $\pm$ 1.54  & 134.6   $\pm$ 23.5 & 3.98   $\pm$ 0.39  & $12$, $19$\\
A754 	       &  0.0535 & 4.31 $\pm$  0.33  & 70.4  $\pm$  23.8  & 1.08   $\pm$ 0.06  & $11$, $2$\\
A773           & 0.2170  & 8.10  $\pm$ 1.35  & 244.3   $\pm$ 31.7 & 1.73  $\pm$  0.17  & $12$, $21$\\
A1300          & 0.3075  & 13.97 $\pm$  2.05 &    --              & 6.09  $\pm$  0.61  & $4$, $15$  \\
A1656 (Coma)   &  0.0231 & 3.77  $\pm$ 0.10  & 154.0   $\pm$ 43.0 & 0.72  $\pm$  0.07  & $11$, $23$, $10$, $25$\\  % K0=154-43+68 (Coma)
A1914          & 0.1712  & 10.71  $\pm$ 1.02 & 63.3   $\pm$ 22.3  & 5.21   $\pm$ 0.24  & $11$, $2$\\
A2163          & 0.2030  & 23.17  $\pm$ 1.48 & 437.3  $\pm$ 82.7  & 18.44   $\pm$ 0.24 & $4$, $16$\\
A2219          & 0.2281  & 12.73  $\pm$ 1.37 &411.6  $\pm$  43.2  & 12.23   $\pm$ 0.59 & $12$, $2$\\
A2254          &0.1780   & 4.32  $\pm$ 0.92  &    --              & 2.94  $\pm$ 0.29   & $11$, $2$\\
A2255 	       &  0.0808 & 2.65  $\pm$ 0.12  & 529.1   $\pm$ 28.2 & 0.89  $\pm$  0.05  & $11$, $22$\\
A2256 	       & 0.0581  & 3.81 $\pm$  0.17  & 349.6   $\pm$ 11.6 & 0.68   $\pm$ 0.12  & $11$, $8$, $6$\\
A2319 	       &  0.0559 & 7.40 $\pm$  0.40  & 270.2  $\pm$   4.8 & 1.12  $\pm$  0.11  & $11$, $15$\\
A2744          & 0.3066  & 12.92 $\pm$  2.41 & 295.1  $\pm$ 113.4 & 17.16   $\pm$ 1.71 & $4$, $21$\\
CL0016+16      & 0.5545  & 18.83  $\pm$ 1.88 &    --              & 6.74  $\pm$  0.67  & $27$, $19$\\
MACSJ0717      & 0.5548  & 24.60  $\pm$ 0.3  & 158.7  $\pm$ 111.6 & 50.00  $\pm$ 10.00 & $14$, $31$, $5$\\
RXCJ2003.5-2323 & 0.3171 & 9.25  $\pm$ 1.53  &    --              & 12.30  $\pm$  0.71 & $4$, $18$\\
\hline
radio mini-halos & & & & & \\
\hline
A426 (Perseus)   & 0.018 &  8.31             & 19.4  $\pm$   0.2 & 4.40             & $1$, $26$\\% Perseus
A2142           & 0.089  & 10.89             & 58.5   $\pm$  2.7 &0.66              & $30$, $19$\\% Giovannini & Feretti 2000: small halo in cool core 
A2390           & 0.2329 & 13.43  $\pm$ 3.16 & 14.7  $\pm$   7.0 & 9.77  $\pm$  0.45& $12$, $2$\\% all mini-halos
A2626           &0.0604  & 1.96              & 23.2   $\pm$  2.9 & 0.43             & $30$, $20$\\
PKS0745-191     & 0.1028 & 14.06             & 11.9   $\pm$  0.7 & 27.00       	    & $30$, $3$\\
RXCJ1314.4-2515 & 0.2439 & 10.94  $\pm$ 1.81 &    --             & 0.75 $\pm$   0.15& $4$, $28$, $17$\\ 
Z7160           & 0.2578 &  8.41  $\pm$ 2.12 & 18.8   $\pm$  3.2 & 2.19  $\pm$  0.26& $12$, $7$\\
\hline
no radio halo detection & & & & & \\
\hline
A141            & 0.2300 & 5.76  $\pm$ 0.90  & 144.1  $\pm$  31.3 & $<$ 0.36     & $4$, $29$\\
A611            & 0.2880 & 8.86  $\pm$ 2.53  & 124.9  $\pm$  18.6 &$<$  0.40     & $13$, $29$\\
A781            & 0.2984 & 11.29  $\pm$ 2.82 &  --                & $<$ 0.36     & $12$, $29$\\
A1423           & 0.2130 & 6.19  $\pm$ 1.34  & 58.8   $\pm$ 12.6  &  $<$ 0.41    & $12$, $29$\\
A2537           & 0.2966 & 10.17  $\pm$ 1.45 & 106.7  $\pm$  19.6 &   $<$ 0.50   & $4$, $29$\\
A2631           & 0.2779 & 7.57  $\pm$ 1.50  & 308.8  $\pm$  37.4 & $<$ 0.39     & $4$, $29$\\
A2667           & 0.2264 & 13.65  $\pm$ 1.38 & 12.3  $\pm$   4.0  & $<$  0.42    & $4$, $29$\\
A2697           & 0.2320 & 6.88  $\pm$ 0.85  &  --                & $<$ 0.40     & $4$, $29$\\  % all upper limits
A3088           & 0.2537 & 6.95  $\pm$ 1.20  & 32.7   $\pm$  9.5  & $<$ 0.42     & $4$, $29$\\
RXCJ1115.8+0129 &0.3499  &13.58  $\pm$ 2.99  & 14.1  $\pm$   5.1  & $<$ 0.45     & $4$, $29$\\
RXCJ1512.2-2254 & 0.3152 & 0.19  $\pm$ 1.76  &  --                & $<$ 0.63     & $4$, $29$\\
RXJ0027.6+2616  & 0.3649 & 12.29  $\pm$ 3.88 &  --                & $<$ 0.68     & $13$, $29$\\
RXJ1532.9+3021  & 0.3450 & 16.49  $\pm$ 4.50 & 14.3  $\pm$  1.9   &  $<$  0.62   & $12$, $29$\\
RXJ2228.6+2037  & 0.4177 & 19.44  $\pm$ 5.55 &  --                &  $<$ 0.91    & $13$, $29$\\
S780            & 0.2357 &  15.53  $\pm$ 2.80&  --                & $<$ 0.36     & $4$, $29$\\
Z2089           & 0.2347 & 6.79  $\pm$ 1.76  & --                 & $<$ 0.27     & $12$, $29$\\
Z2701           & 0.2140 & 6.59 $\pm$  1.15  & 34.0   $\pm$   4.2 &  $<$ 0.42    & $12$, $29$ \\
Z5699           & 0.3063 & 8.96  $\pm$ 2.24  &   --               &  $<$ 0.54    & $13$, $29$\\
Z5768           & 0.2660 & 7.47  $\pm$ 1.66  &  --                &  $<$ 0.36    & $13$, $29$\\
Z7215           & 0.2897 & 7.34  $\pm$ 1.91  &   --               &  $<$ 0.55    & $13$, $29$\\
\hline
\end{tabular}
\end{center}

%\vspace{-2em} %%  only to match referee format
 \caption{Cluster sample with radio halo detections and upper limits.
   Sample base from \citet{2009A&A...507..661B}.  Four mini-halos are
   added from \citet{2004A&A...417....1G}.  The X-ray luminosities are
   as in \citet{2009A&A...507..661B}, for the four additional
   mini-halos data was added from \citet{2002ApJ...567..716R}, and for
   A2626 from \citet{2008MNRAS.384.1502S}.  Central values for the
   entropy indicator $K_0=kT_\rmn{x,0}^{} n_\rmn{e,0}^{-2/3}$ are taken
   from the extrapolation method in \citet{2009ApJS..182...12C}
   applied to Chandra data. $K_0$ of Coma at 12 kpc is from
   \citet{2008ApJ...687..899R}.
%   \newline %% only to match referee format
   References: 1 = \citet{ALLEN1992MNRAS.254...51A}, 2 = Bacchi
   et~al. (\citeyear{Bacchi2003A&A...400..465B}), 3 =
   \citet{BAUMODEA1991MNRAS.250..737B}, 4 = B{\"o}hringer
   et~al. (\citeyear{Boehringer2004A&A...425..367B}), 5 = Bonafede
   et~al. (\citeyear{Bonafede2009A&A...503..707B}), 6 = Brentjens
   (\citeyear{Brentjens2008A&A...489...69B}), 7 = Cassano
   et~al. (\citeyear{2008A&A...480..687C}), 8 = Clarke \& En{\ss}lin
   (\citeyear{2006AJ....131.2900C}), 9 = Dallacasa
   et~al. (\citeyear{2009ApJ...699.1288D}), 10 = Deiss
   et~al. (\citeyear{1997A&A...321...55D}), 11 = Ebeling
   et~al. (\citeyear{Ebeling1996MNRAS.281..799E}), 12 = Ebeling
   et~al. (\citeyear{Ebeling1998MNRAS.301..881E}), 13 = Ebeling
   et~al. (\citeyear{Ebeling2000MNRAS.318..333E}), 14 = Ebeling
   et~al. (\citeyear{Ebeling2007ApJ...661L..33E}), 15 = Feretti
   (\citeyear{Feretti2002IAUS..199..133F}), 16 = Feretti
   et~al. (\citeyear{Feretti2001A&A...373..106F}), 17 = Giacintucci
   (\citeyear{GiacintucciPhD}), 18 = Giacintucci
   et~al. (\citeyear{2009A&A...505...45G}), 19 = Giovannini \& Feretti
   (\citeyear{2000NewA....5..335G}), 20 = \citet{2004A&A...417....1G},
   21 = Govoni et~al. (\citeyear{Govoni2001A&A...376..803G}), 22 =
   Govoni et~al. (\citeyear{Govoni2005A&A...430L...5G}), 23 = Kim
   et~al. (\citeyear{Kim1990ApJ...355...29K}), 24 = Liang
   et~al. (\citeyear{2000ApJ...544..686L}), 25 = Rafferty et~al. 
   (\citeyear{2008ApJ...687..899R}), 26 = Sijbring (\citeyear{SijbringPhD}),
   27 = Tsuru et~al. (\citeyear{Tsuru1996uxsa.conf..375T}), 28 =
   Venturi et~al. (\citeyear{Venturi2007A&A...463..937V}), 29 =
   \citet{2008A&A...484..327V}, 30 = \citet{WHITE1997MNRAS.292..419W},
   31 = van Weeren et~al. (\citeyear{vanWeeren2009A&A...505..991V}).
 }
%\end{singlespace} %%  only to match referee format
\label{tab:clusterstat}
\end{table*}

Cluster radio halos are our primary evidence for the existence of CR
in galaxy clusters. They are spatially extended regions of diffuse
radio emission, which have regular morphologies, very much like the
morphology of the X-ray emitting thermal ICM plasma. Their radio
synchrotron emission is unpolarised, due to the contribution of
various magnetic field orientations along the line of sight, and
Faraday rotation de-polarisation. A compilation of radio halo
luminosities and X-ray properties of their hosting clusters can be
found in Tab.~\ref{tab:clusterstat} and Fig.~\ref{fig:radiohalostat}.

Cluster radio halos have to be discriminated from cluster radio relics \citep[for taxonomy see][]{2004rcfg.procE..25K}. 
Relics are also extended, but often located at the cluster periphery of galaxy clusters. 
They show an irregular morphology and are often significantly polarised. As halos, they appear preferentially in clusters exhibiting signs of merging activity. 
A halo and relic can appear in the same cluster\footnote{e.g. A2256: \citet{1976A&A....52..107B, 1978MNRAS.185..607M, 1979A&A....80..201B, 1994ApJ...436..654R, 2006AJ....131.2900C, 2010ApJ...718..939K}}, or even two relics simultaneously have been observed\footnote{e.g. A3667 and A3376: \citet{1975MmRAS..79....1S, 1982MNRAS.198..259G, 1992ApJS...80..137J, 1997MNRAS.290..577R, 2006Sci...314..791B}}.
 Relics are believed to trace merger and accretion shock waves, either due to direct particle acceleration via the Fermi-I mechanism\footnote{Radio gischt in the taxonomy of \citet{2004rcfg.procE..25K}: \citet{1998A&A...332..395E, 1999ApJ...518..603R,2001ApJ...559..785K, 2001ApJ...562..233M,  2002NewA....7..249B, 2003ApJ...594..709B, 2004MNRAS.347..389H, 2008MNRAS.391.1511H, 2004AcPPB..35.2131S, 2004NewAR..48.1119K, 2007MNRAS.375...77H, 2008A&A...486..347G,  2009MNRAS.393.1073B}}, or due to the revival of old radio cocoons of radio galaxies via compression\footnote{Radio phoenix in the taxonomy of \citet{2004rcfg.procE..25K}: \citet{2001A&A...366...26E, 2001AJ....122.1172S, 2001ApJ...549L..39E, 2002MNRAS.331.1011E, 2002MNRAS.336..649K, 2003astro.ph..9612B, 2004MPLA...19.2317G}}. Multi relic formation emerges naturally in numerical models of structure formation shock accelerated CR electrons~\citep{1999ApJ...518..603R, 2001ApJ...562..233M}.

Radio halos and relics are preferentially found in clusters showing signs of merger activity\footnote{\citet{2001ApJ...553L..15B, 2001A&A...378..408S, 2001ApJ...563...95M, 2004ApJ...605..695G, 2008A&A...484..327V, 2010arXiv1008.3624C}}.
Radio halos are a larger mystery than relics. Since the radiating CRe are short lived (0.1 Gyr, see Fig. \ref{fig:ecooltime}) they have to be replenished or re-accelerated in-situ. Two classes of models are currently under discussion, the hadronic and the re-acceleration model. A combination or coexistence of their underlying processes might be in operation in clusters, as the re-accelerated CRe could be injected by hadronic CRp-gas interactions \citep{2005MNRAS.363.1173B}. However, this requires some fine-tuning of parameters%
\footnote{The argumentation goes as following: If secondary CRe from hadronically interacting CRp get an energy boost by some factor by re-acceleration in order to produce the observed radio halo, a similar boost (or even larger) can be expected for the CRp. However, such a boost of the CRp population would increase the amount of secondary CRe to the level required to explain the radio halo directly. Only if some plasma physical reasons can be found why CRe are more efficiently accelerated than CRp (despite the higher losses of the former), such a hybrid scenario can be expected to operate.} 
and it is more likely that one process is responsible for most of the radio halo emission and the other process is subdominant.
Alternatively, different regions of the same halo could be generated by the two different mechanisms.  \citet{2008MNRAS.385.1211P} proposes that the central part of cluster radio halos is hadronic due to the high target density there, whereas at the outskirts shock waves have higher Mach numbers and can provide Fermi~I acceleration.

Cluster radio halos come in two sizes: cluster wide and therefore giant radio halos and radio mini-halos. The former are predominantly found in clusters showing merger activities whereas the latter are found in very relaxed clusters which developed a cool core, which harbors the mini-halo. The radio luminosity of giant halos seems to be strongly correlated with the X-ray emissivity of the cluster \citep[][ and see Fig. \ref{fig:radiohalostat}]{2000ApJ...544..686L,2009A&A...507..661B}. The radio luminosities of the mini-halos also seem to correlate in the same way with the cluster X-ray luminosity, which itself is usually dominated by the cool core emission.  

A large fraction of clusters do not exhibit significant radio halo emission of any kind, and only upper limits to their synchrotron flux are known. About half of the radio deficient clusters, for which we have Chandra data, show clear evidence for some level of cool core structure ($K_0 \lesssim 50\,\rmn{keV\,cm^2}$) as can be seen in Fig. \ref{fig:radiohalostat}. This could either imply that these clusters are in the intermediate state between having giant radio halos because of merging activity and having mini halos due to strongly developed cool cores. On the other hand there could be two populations of clusters -- cool cores and non-cool cores -- and the corresponding radio luminosity responds sensitively to the level of injected turbulence by either AGN or cluster mergers, respectively.

Therefore, clusters at the same X-ray luminosity seem to be bimodal with respect to their radio luminosity. 
Either they have a prominent halo or they do not exhibit any detectable diffuse radio halo emission. 
This indicates the existence of pronounced and rapidly operating switch-on/switch-off mechanisms, which are able to change the radio luminosity by at least a factor of $10-30$ \citep{2009A&A...507..661B}. 
Thus a mechanism is easily realised in the re-acceleration model of halo formation, due to the short cooling time of the radio emitting electrons, which just cool away once turbulence is unable to maintain them. 
It is less easily realised in the hadronic model via magnetic field decay after turbulence as proposed by \citet{2009JCAP...09..024K} and \citet{2010arXiv1003.1133K}, since the turbulent decay takes about a Gyr and the magnetic field decay is relatively gentle \citep{2006MNRAS.366.1437S}. In addition, this argument appears weak because
$\mu$G strong magnetic fields are commonly observed in clusters without diffuse radio emission~\citep{2001ApJ...547L.111C}.

However, the assumption that clusters evolve only vertically on the $L_\nu-L_\rmn{X}$ plane, as it is the basis of the above argumentation on bimodality, is probably incorrect. The central entropy of many of the radio halo deficient clusters is low, indicating that they host or are forming a cool core. Cool cores tend to dominate the X-ray luminosity of clusters. Therefore, also a strong horizontal evolution in this plane can be expected once the radio halo luminosity decreases as the cluster relaxes after a merger. Radiative cooling in the central cluster regions increases the gas density and therefore the X-ray luminosity while it also decreases $K_0$. This alleviates the requirements on the speed and the magnitude a radio halo switch mechanism must fulfill in order to explain the observations.

Nevertheless, such a mechanism is probably needed to understand the observational data. It has to full-fill two requirements: it should be able to extinguish a radio halo, but also not prevent it from being switched on again later, otherwise we would not observe halos in the present universe. The mechanism we are discussing in this work is actually able to switch radio halos on and off in both models, the hadronic and re-acceleration model. Since both have their individual strengths and weaknesses in explaining the different observational features of radio halos, we want to briefly discuss those first.

\subsection{Hadronic models}
In the hadronic model the accumulated CRp inject continuously the radio emitting CRe into the ICM due to well known hadronic process $p_\mathrm{CR} + p \rightarrow \pi^{\pm} + \ldots \rightarrow e^{\pm} + \nu_\rmn{e}/\bar{\nu}_\rmn{e} + \nu_\mu+\bar{\nu}_\mu + \dots$, (see  footnote \ref{note:hadronic} for references). 
The CRe loose their energy nearly instantaneously (see Fig.~\ref{fig:ecooltime}) and therefore the radio emission traces a combination of the CRp population, the distribution of the target gas density, and the magnetic field profile of the galaxy cluster in the case of weak magnetic field strengths  $B<B_\rmn{cmb} = 3.27 \, (1+z)^2 \, \mu$G ($z$ being the redshift) and without magnetic field dependence in the strong-field case.

The hadronic model has a number of \textbf{advantages}:
\begin{itemize}
 \item Since CRp are expected to be present in the ICM due to structure formation shock waves, active galactic nuclei, and the deposition of interstellar media of galaxies, radio halos emerge very naturally in these models. 
 \item Also the observed power-law spectra of many radio halos \citep[e.g. in the Bullet cluster,][]{2000ApJ...544..686L} emerge naturally, since they reflect the power-law spectra of CRp acceleration.
 \item The observed smooth, regular morphology of halos is also expected in hadronic models, since the long lived CRp had enough time to become distributed within the cluster volume \citep{2001ApJ...559...59M,2003MNRAS.342.1009M,2008MNRAS.385.1211P}.
 \item The observed correlation of radio halo luminosity with X-ray luminosity ($L_\nu-L_X$ relation, see Fig. \ref{fig:radiohalostat}) of clusters with halos holds within these models for plausible magnetic field values and CRp acceleration efficiencies \citep{2000A&A...362..151D,2001ApJ...562..233M,2008MNRAS.385.1242P, 2010arXiv1003.1133K}. 
 \item  Also the correlation of radio halo surface brightness with X-ray surface brightness of clusters with halos holds roughly for sensible magnetic field profiles and CRp acceleration efficiencies. [However, \citet{2001A&A...369..441G} find that the naively expected radio halo profile does not seem to fit the observed asymptotic in case of weak magnetic fields ($B<3 \mu$G).]
\end{itemize}

However, there are also a number of \textbf{issues} with the hadronic model:
\begin{itemize}
 \item About two thirds of the most X-ray luminous clusters do not exhibit radio halos, whereas the hadronic model seems to suggests that all clusters exhibit halos \citep{2001ApJ...562..233M, 2003MNRAS.342.1009M,  2008MNRAS.385.1242P, 2010arXiv1001.5023P}. [This problem will be addressed and alleviated by this work.]
 \item The curvature in the total spectrum claimed for the Coma cluster radio halo \citep{2001MNRAS.320..365B} are not reproduced in current numerical models of the hadronic scenario. 
[However, the particle transport in these models 
neglects diffusion and streaming which is potentially important. In addition, the high frequency observations of the steep spectrum radio halo, on which this claim rests, are extremely difficult, since point sources much brighter than the diffuse radio halo itself have to be subtracted, and the sensitivity to large scale emission of radio interferometers is reduced at higher frequencies. Thus the curvature might be partly a fluke, as also some inconsistencies of flux measurements at the same frequencies in this spectrum indicate \citep[see 0.4 and 1.4 GHz fluxes reported in][]{2003A&A...397...53T}.]
 \item Spectral steepening at the edges as proposed by some observations \citep{2001MNRAS.320..365B, 2004JKAS...37..315F} can not be explained. [However, the observational arguments above are even more severe for this. In addition, the negative flux of the Sunyaev-Zel'dovich decrement of clusters affects the spectrum especially hard in the outskirts, and also provides some bending to the total spectrum \citep{2002A&A...396L..17E, 2004MNRAS.352...76P}. Finally, energy dependent CRp transport would create such spectral variations, as argued in this work.]
\end{itemize}

The hadronic model makes one hard prediction, which hopefully will permit its confirmation or rebuttal at some point in the future. The radio halo emission should always be accompanied by some level of gamma-ray flux, due to the hadronic production of neutral pions and their decay into gamma-rays, $p_\mathrm{CR} + p \rightarrow \pi^{0} + \ldots\rightarrow 2\, \gamma+ \ldots$ The current upper limits on diffuse gamma-ray flux from cluster of galaxies by the Fermi collaboration \citep{2010ApJ...717L..71A} are still well above the predictions of expected fluxes, even for the most optimistic assumptions about the CR acceleration efficiency \citep{2010arXiv1001.5023P} or by tying the expected gamma-ray emission to the simulated radio halo emission for reasonable assumed magnetic field strengths \citep{ 2010MNRAS.401...47D, 2010arXiv1003.0336D}. They are far off the minimal gamma-ray flux expected in the limit of strong magnetic field strength \citep[$\gg 3 \mu$G;][]{2008MNRAS.385.1242P, 2010ApJ...710..634A}. 
 \citet{2009A&A...508..599B} and \citet{2010arXiv1006.1648J} argue that unusually strong ICM magnetic fields would be required by the hadronic model  for radio halos with reported steep radio spectra (e.g. Abell 1914, 2256). 
However, this is only true if the CRp spectra can be extrapolated from the radio emitting energies ($\sim$100 GeV) into the subrelativistic regime ($<1$~GeV) without any spectral break as assumed by these authors.

\subsection{Re-acceleration models}
In re-acceleration models, a pre-existing CRe population at lower energies of about 0.1-10 GeV gets re-accelerated  into the radio emitting regime of about 10 GeV by plasma waves (see references in footnote \ref{note:reacc}). 
These are generated by the turbulence during and after a cluster merger event. 
Some level of re-acceleration has to happen most of the time or frequently enough in order to prevent the CRe population in the cluster center from loosing its energy completely due to Coulomb losses on a timescale of about 1 Gyr.

Also the re-acceleration model has its \textbf{advantages}:
\begin{itemize}
 \item The bimodality of radio halo luminosities is explained in this model by the presence and decay of the re-accelerating turbulence in merging and relaxed clusters, respectively \citep{2009A&A...507..661B}.
 \item CRe are expected to be accumulated in the ICM due to injection by radio galaxy outflows \citep{1993ApJ...406..399G} and acceleration at shocks, if some level of continuous re-acceleration can prevent them from thermalisation.
 \item Diffusive re-acceleration is a natural plasma process, which must occur in turbulent astrophysical environments as clusters \citep{1987A&A...182...21S}.
 \item The complex morphologies reported for some radio halos come naturally about due to the effect of intermittency of turbulence \citep{2004JKAS...37..315F}.
 \item Significantly curved radio spectra are also very natural, since the interplay of acceleration and cooling produces spectral bumps and cut-offs  \citep{2004JKAS...37..315F, 2007MNRAS.378..245B, 2010arXiv1008.0184B}.
 \item The $L_\nu-L_X$ relation holds also for reasonable magnetic fields \citep{2007MNRAS.378.1565C}.
\end{itemize}

\begin{figure*}
 \includegraphics[bb=135 242 478 544,width =0.49\textwidth]{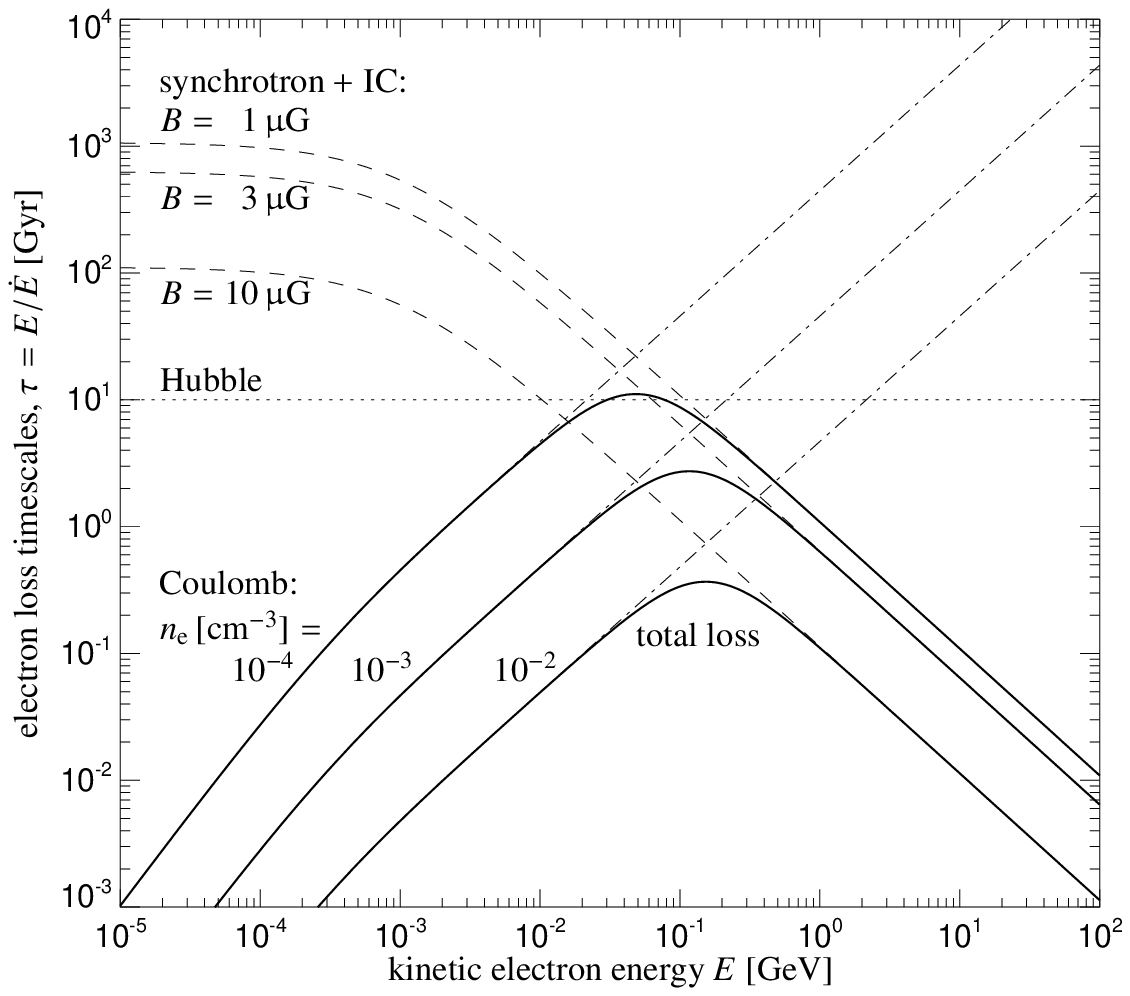}
 % timescales.eps: 0x0 pixel, 300dpi, 0.00x0.00 cm, bb=135 242 478 544
\includegraphics[bb=135 242 478 544,width =0.49\textwidth]{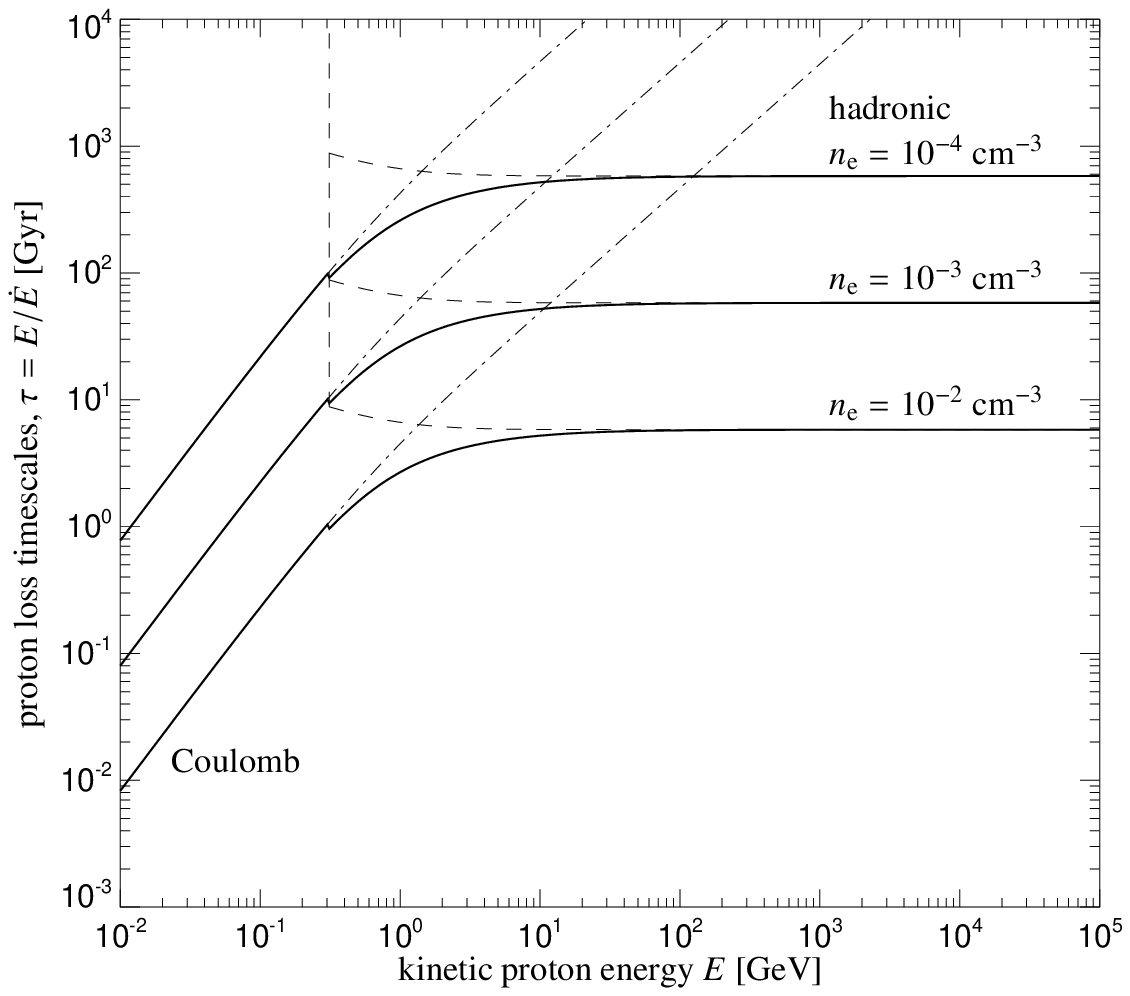}
% timescales_proton.eps: 0x0 pixel, 300dpi, 0.00x0.00 cm, bb=135 242 478 544
 \caption{Cooling time of CR in the ICM as a function of their kinetic energy. \textbf{Left:} CRe cooling times for typical densities and magnetic field strength that range from the central to the peripheral regions in galaxy clusters. Coulomb and IC/synchrotron cooling is modeled following \citet{1972Phy....60..145G} and \citet{1979rpa..book.....R}, respectively. \textbf{Right:} CRp cooling times for the same densities. Coulomb cooling is modeled following \citet{1972Phy....58..379G}. The hadronic cooling time above the kinematic threshold for pion production is $\tau_\rmn{pp} = 1/(0.5\,\sigma_\rmn{pp}\, n_\rmn{N}\, \vel_\rmn{CR})$ with  $\sigma_\rmn{pp} = 32~$mbarn, an inelasticity of $\sim 50\%$ and $\vel_\rmn{CR}$ the CRp velocity. It is apparent that CRp above 10 GeV have livetimes in the ICM at least 60 times longer than CRe at any energy. CRe can survive for a Hubble time without re-acceleration only within the dilute outskirts of clusters. The radio emitting electrons have an energy of about $10~$GeV in $\mu$G fields, and herefore a lifetime of 0.1 Gyr or less. If they are of hadronic origin, their parent CRp had energies of about 100 GeV, which have considerable longer lifetimes. 
}
 \label{fig:ecooltime}
\end{figure*}

The \textbf{issues} with the re-acceleration model are:
\begin{itemize}
 \item Second order Fermi acceleration is known to be very inefficient since the efficiency scales with $(\vel_\rmn{wave}/c)^2\ll1$. The acceleration efficiencies used in the re-acceleration models are difficult to be derived from first principles and are often fit to reproduce the observations (e.g. \citet{2001MNRAS.320..365B} adopt the magnetic field profile to match their re-acceleration profile). [However, magnetosonic turbulence may be more efficient, and could provide sufficient re-acceleration if the compressible turbulence is of the required strength.
\citet{2007MNRAS.378..245B,2010arXiv1008.0184B} show that a compressible wavefield of 15-30\,\% of the thermal energy content of the cluster, reaching down to small scales with a Kraichnan-scaling, would be sufficient. It should be noted here that Fermi~I acceleration at ICM shock waves is also a promising re-acceleration mechanism.]
\item The maximal cooling time of CRe is $\tau_\rmn{e}\sim 1\,\rmn{Gyr}\,(n_\rmn{e}/(3\times10^{-3}\,\rmn{cm}^{-3}))$ (see Fig. \ref{fig:ecooltime}). 
Without replenishing the 100 MeV-seed population of CRe, the re-acceleration would not be able to generate 10 GeV CRe that radiate GHz radio waves. 
 Either continuous injection or re-acceleration of the short-lived central CRe population is needed to counteract cooling or a process that moves CRe inwards without much losses from the dilute outer cluster regions where the cooling time approaches the Hubble time. [This problem will be addressed and alleviated by this work.]
\item Any re-acceleration mechanism is unable to discriminate between CRe and CRp at GeV energies. 
In case re-acceleration operates, the CRp population gains more than the CRe population in the long run, due to the much lower losses. 
Given that the injection efficiency of CRp by shock waves is expected to be hundred times higher than the CRe one, the re-accelerated CRe can easily be outnumbered by hadronically injected CRe. [However, the possibility exists that the CR population is dominated by injection of radio plasma, which might contain mostly CRe.]
\item The natural curvature of re-acceleration radio spectra requires fine tuning to reproduce the observed power-law radio spectra seen in some radio halos. [However, a detailed statistics of the radio halo spectral slopes that includes statistical and systematic uncertainties is not published yet.]
\item The regular morphology of many radio halos might also be a challenge to the re-acceleration model, since turbulence and therefore re-acceleration is expected to be intermittent. [However, detailed numerical simulations of the expected morphologies are not yet published].
\end{itemize}

The hardest prediction of the re-acceleration model so far is that low X-ray luminous clusters should not exhibit radio halos \citep{2006MNRAS.369.1577C,2008A&A...480..687C}. This prediction will become testable with upcoming sensitive radio telescope arrays.

\section{Cosmic ray transport in galaxy clusters }
\label{sec:CRtransp}
\subsection{Cluster weather conditions}

The atmospheric conditions in galaxy clusters differ strongly depending on the time elapsed since the last cluster merger event. During a cluster merger, bulk velocities close and above the sound speed of $c_\mathrm{s} \approx$ 1000-2000 km/s are injected on scales of a few $\times 100$ kpc.
A fair fraction of this energy goes into turbulent motions, which can account for 10-20\% of the thermal pressure of the ICM~\citep{2004A&A...426..387S,2010arXiv1001.1170P}.
A good fraction, however, is directly dissipated in shock waves, mostly in the cluster center, however with the highest Mach number shock waves and therefore the most efficient CR acceleration sites appearing in the cluster outskirts \citep{2000ApJ...542..608M,2006MNRAS.367..113P, 2008MNRAS.385.1211P}.

The cluster turbulence persists for about a Gyr, partly because the turbulent decay of eddies takes several eddy turn over times to transport the kinetic energy to dissipative scales, partly because the gravitational drag of the merging dark matter halos continues to stir turbulence for some time\footnote{The following papers provide insight into cluster turbulence: \citet{1998ApJ...495...80B,1999ApJ...518..594R,  2003AstL...29..791I, 2004A&A...426..387S, 2005MNRAS.358..139F, 2005MNRAS.364..753D, 2006MNRAS.366.1437S,  2006PhPl...13e6501S, 2006MNRAS.369L..14V, 2008MNRAS.388.1079I, 2008MNRAS.388.1089I, 2009ApJ...707...40M, 2009MNRAS.393.1073B, 2009A&A...504...33V, 2009arXiv0912.3930K, 2010arXiv1001.1170P, 2010arXiv1010.4492S}.}.

As the turbulence decays, the cluster atmosphere settles into a stratification with the lowest entropy gas at the bottom and higher entropy gas as larger radii. The turbulence may become more two-dimensional, since all radial motions are working against gravitational and/or pressure forces. If the cluster becomes sufficiently quiet, thermal instabilities might set in due to heat flux along magnetic field lines, rearranging the field in preferentially radial orientation as predicted by the magneto-thermal instability \citep{2000ApJ...534..420B, 2007ApJ...664..135P, 2008ApJ...688..905P, 2010NatPh...6..520P}.

At the dense cluster center, X-ray cooling could become catastrophic due to the onset of a cooling instability. However, the developing cool cores seem to be stabilised against a complete collapse by energetic feedback from the central galaxy in form of AGN outflows \citep[e.g.][]{2002MNRAS.332..729C} or supernovae driven winds, both fed by condensing gas from the cooling region. Since cool cores have temperatures of only a few keV, their sound speed is below 1000 km/s. The turbulence, which is most likely subsonic there, will therefore only have a speed of a few 100 km/s there \citep{2003MNRAS.344L..48F, 2006A&A...453..447E}.

Let us model the cluster weather in a simplistic fashion, in order to estimate the order of magnitude of the transport processes. We assume that the turbulence is injected on a length scale $L_\mathrm{tu}$ comparable to the cluster core radius $r_\rmn{c} \sim 200$ kpc, which is also roughly the atmospheric scale height in the inner part of the cluster:
\begin{equation}
 L_\mathrm{tu} = \chi_\mathrm{tu}\, r_\rmn{c} \;\;\; \mbox{with} \;\;\;\chi_\mathrm{tu} \sim 1.
\end{equation}
Although the turbulence injection is supersonic, shock waves rapidly dissipate kinetic energy until turbulent velocities $\vel_\mathrm{tu}$ become subsonic. Thus, the turbulence stays nearly trans-sonic for about a Gyr. We therefore write
\begin{equation}
 \vel_\mathrm{tu} = \alpha_\mathrm{tu}\, c_\mathrm{s}\;\;\; \mbox{with} \;\;\; \alpha_\mathrm{tu}(t) \lesssim 1
\end{equation}
clearly being time dependent. Note that for notational consistency reasons, we identify the turbulent Mach number by $M_\rmn{tu}\equiv \alpha_\rmn{tu}$. Half an eddy turn over time
\begin{equation}
 \tau_\mathrm{tu} = \frac{\pi}{2}\, \frac{L_\mathrm{tu}}{\vel_\mathrm{tu}} = \frac{\pi}{2}\,  \frac{\tau_\mathrm{s}}{\alpha_\mathrm{tu}} 
\end{equation}
is probably larger, but comparable to the sound crossing time of an eddy $\tau_\mathrm{s} =  L_\mathrm{tu} / c_\mathrm{s}$.

The hydrodynamical turbulence is expected to drive a small-scale magnetic dynamo, which is believed to saturate for magnetic energies, which are a fraction $\alpha_B^2 \approx 0.03-1$ of the kinetic energy density, and a magnetic coherence length $\lambda_B$ which is smaller than the turbulent injection scale by some factor $\mathcal{O}(1-10)$ \citep{2006MNRAS.366.1437S, 2006A&A...453..447E}. We have to stress that for filamentary fields, the perpendicular scale of the magnetic structures, which dominates this number, is much smaller than the parallel scale. Flux tubes can have extension $L_B$ comparable to the eddy size.
\footnote{
During the kinematic stage of the small-scale or fluctuation dynamo,
the perpendicular scale is the microscopic resistive scale.
How this changes due to nonlinear effects is not 
yet fully clear (For papers on this aspect see 
\citet{2004PhRvE..70a6308H, 2004ApJ...612..276S,
 2005PhR...417....1B}). An increase in 
the effective resistivity due to nonlinear effects \citep{1999PhRvL..83.2957S}
or due to microscopic plasma effects \citep{2006PhPl...13e6501S} 
is needed if the perpendicular
scales have to become a fraction of the turbulent scale,
as required to explain the observed coherence of 
cluster magnetic fields \citep{2006A&A...453..447E}.
}

We parametrise the Alfv\'enic velocity by 
\begin{equation}
 \vel_\mathrm{A} = \alpha_B\, \vel_\mathrm{tu} = \alpha_B\,\alpha_\mathrm{tu}\, c_\mathrm{s}
\end{equation}
and the magnetic bending-scale by
\begin{equation}
 L_{B} = \chi_{B}\, L_\mathrm{tu} \;\;\; \mbox{with} \;\;\;\chi_{B} \sim 1.
\end{equation}
The ICM magnetic field strength is a relatively poorly known number, despite substantial efforts to measure it in the different cluster environments. Inverse Compton based methods\footnote{e.g. \citet{1994ApJ...429..554R, 1998A&A...330...90E, 1998ApJ...494L.177S, 1998ApJ...506..502B, 1999APh....11...73V, 1999ApJ...520..529S, 2000ApJ...533...73S, 2000ApJ...535...45A}} suggest rather low field strength of $\mathcal{O}(0.3 \, \mu\mathrm{G})$. These methods identify reported excess emission above the thermal ICM X-rays in the extreme ultraviolet and hard X-ray bands with CMB photons  up-scattered by the radio halo emitting electrons. However, the identification is not unique, and even the detection of the excess emission is heavily debated \citep[][ and references therein]{2008xru..confE.249M}. Recent work using the Swift/Burst Alert Telescope \citep{2009ApJ...690..367A, 2010arXiv1009.4699A} was able to reject the hypothesis of non-thermal power-law emission in a sample of 20 clusters, except for the Bullet cluster where they seem to require a non-thermal power-law component. They are able to obtain excellent fits of their data with multi-temperature models  -- hence inferred values of the magnetic field strength using the inverse Compton method are better regarded as lower limits on the field strength. Magnetic field measurements using Faraday rotation data of galaxy clusters \citep[e.g.][]{2001ApJ...547L.111C} find fields $\mathcal{O}(3\,\mu\mathrm{G})$, however, these numbers depend on the assumed characteristic field scale $\lambda_B$. Note that the IC-based magnetic field strengths are weighted by the CRe density at $\sim 10$ GeV while the Faraday rotation-based values are weighted by the thermal electron density. In the case of the Hydra A cool core, where high quality data enables  measuring the magnetic power spectrum in detail, magnetic fields of $\mathcal{O}(10-30\,\mu\mathrm{G})$ seem to support the idea of a saturated field strength with $\alpha_B^2 \sim \mathcal{O}(1)$ \citep{2009arXiv0912.3930K}.

Anyhow, it will turn out that the precise magnetic field strength is less important for our argumentation than the characteristic bending length and the assumption that the magnetic energy density is proportional to the turbulent one, irrespective the numerical factor in between them.

\subsection{CR transport}

The discussion in this section is applicable for CR electrons as well as protons. We describe our CR population by their phase space density $f(\vec{r},p,t) = d^4N/(d\vec{r}\,d p)$ assuming that the momentum distribution is nearly isotropic and can hence be described by a one-dimensional spectrum. Anisotropies lead to CR streaming with a velocity $\vel_\mathrm{cr}\sim c_\mathrm{s}$, as we will argue below. Therefore anisotropies are expected only on the order $\vel_\mathrm{cr}/c_\mathrm{cr} \sim c_\mathrm{s}/c \sim  10^{-3}$ which we can safely neglect here except for their transport effect. Here $c_\mathrm{cr} \lesssim c$ denotes the CR particle velocity, which is close to the speed of light $c$. We express the CR momentum $p$ in units of $m\,c$, with $m$ being the particle mass. For CRp, we assume a power-law particle spectrum,
\begin{equation} \label{eq:CRspec}
f(\vec{r},p,t) = C(\vec{r},t)\, p^{-\alpha}.
\end{equation}
Here $\alpha(\vec{r}, t) \approx \alpha \approx 2.1-2.5$ is the dominating CRp spectral index around 100 GeV expected in the ICM from structure formation shock waves \citep{2001ApJ...559...59M,2010arXiv1001.5023P} and $C(\vec{r},t)$ the spectral normalisation constant, which will capture the spatial variations and temporal evolution.

The full CR transport equation reads
\begin{equation}
\frac{\partial f}{\partial t} + \frac{\partial }{\partial r_i} (\vel_i \,f)  + \frac{\partial }{\partial p} (\dot{p} \,f) = \frac{\partial }{\partial {r}_i} \left( \kappa_{ij} \frac{\partial }{\partial {r}_j} f \right)+ q - \frac{f}{\tau_\mathrm{loss}}.
\end{equation}
Here, $\vvel = \vvel_\mathrm{tu}+\vvel_\mathrm{st}$ is the CR transport velocity, which is the sum of the (turbulent) ICM gas velocity, $\vvel_\mathrm{tu}$, and the CR streaming velocity with respect to the gas $\vvel_\mathrm{st}$, to be discussed below. For relativistic protons, the continuous momentum loss or gain term $\dot{p}(\vec{r},p, t) $ can be approximated to be solely due to adiabatic energy changes, since Coulomb and radiative losses can be ignored and hadronic losses are modelled as catastrophic losses with a timescale $\tau_\mathrm{loss} \sim 1/(\sigma_\rmn{pp}\,n_\mathrm{p}\,c) \sim \mathcal{O}(10 \,\mbox{Gyr})$ (see Fig. \ref{fig:ecooltime}). Thus
\begin{equation}
 \dot{p} \approx \dot{p}^\mathrm{ad} = - \frac{p}{3}  \, \vec{\nabla}\cdot \vvel.
\end{equation}
The adiabatic energy changes of CRe follow the same equation, however, radiative losses and Coulomb losses can usually not be ignored for CRe (see Fig. \ref{fig:ecooltime}). Since we do not want to enter a detailed modeling of the CRe spectrum for the sake of briefness, we ignore the cooling in our calculations and can therefore only make timescale arguments just to see if and how CRe transport processes might be relevant.

The total transport velocity $\vvel = \vvel_\mathrm{tu}+\vvel_\mathrm{st}$ appears in the adiabatic term, since $\vvel$ represents the combined frame of the plasma waves and magnetic field line motions which together confine the particles a bit like {\it walls}. Thus, the energy transfer from streaming CR into waves is an adiabatic process for the CR, the particles push these  {\it walls} away while expanding into unoccupied regions. The energy deposited in the waves gets dissipated by Landau and other damping processes and the interplay of excitation by the CR streaming and damping determines the precise value of the streaming velocity $\vvel_\mathrm{st}$. This will be discussed more in detail in Sect. \ref{sec:CRstreaming}.
For the time being we anticipate the result adopted here that
\begin{equation}\label{vst::eq}
 \vel_\mathrm{st} = \alpha_\mathrm{st}\, c_\mathrm{s} \;\;\; \mbox{with}\;\;\; \alpha_\mathrm{st} \lesssim 1.
\end{equation}
In addition, we will assume that the diffusive transport (discussed below) is possibly less efficient,
since if it leads to significant CR bulk motions, these will excite
plasma waves which limits the transport to the streaming regime.

The CR diffusion tensor is highly anisotropic with $\kappa_\| $ and $\kappa_\perp$ being the diffusion coefficients parallel and perpendicular to the local magnetic field direction, respectively. Nondiagonal terms that mix the parallel and perpendicular direction appear in case of diverging or converging magnetic flux. Very often, $\kappa_\|  \gg \kappa_\perp$ so that we can ignore cross field diffusion on short timescales. Although several theoretical works%
\footnote{\citet{1966ApJ...146..480J,1967ApJ...149..405J,1978PhRvL..40...38R,1995A&A...302L..21D,1997ApJ...485..655B,1997A&A...326..793M,1998A&A...337..558M,1999ApJ...520..204G,2001ApJ...562L.129N}}
 claim that in turbulent environments $\kappa_\perp$ might become comparable to $\kappa_\|$, the observation of sharply edged boundaries of radio plasma questions if the necessary conditions for this are realised in the ICM.

When the CR are tied to their flux tube it makes sense to introduce an affine coordinate $x$ along the tube and a CR density per infinitesimal magnetic flux $d^2\phi = B\,dy\,dz$ with $g(x,p,t) = d^4N/(d^2\phi \,dx \, dp)$. The transport equation along the flux tube can be written as 
\begin{equation}
\frac{\partial g}{\partial t} + \frac{\partial }{\partial x} (\vel_x \,g)  + \frac{\partial }{\partial p} (\dot{p} \,g) = \frac{\partial }{\partial {x}} \left( \frac{\kappa_{\|}}{B} \frac{\partial }{\partial {x}} (B\, g) \right)
\end{equation}
\citep{1978A&A....70..367C, 2003A&A...399..409E},
where we dropped the injection term $q$ and the catastrophic losses. The magnetic field enters the transport equation since magnetic mirrors reflect particles away from locations with converging field lines, so that the CR population tries to establish a constant density per volume, and not per magnetic flux. 
Regions where the magnetic field strength is lower and therefore the flux tube diameter is larger will carry more CRs per tube length. 
Since for a flux tube which connects volume elements on the bottom and top of an atmospheric  scale height (or the fraction $\chi_B\, \chi_\mathrm{tu}$ of it) the tube diameter is expected to be larger at larger cluster radii. This is due to the low-pressure environment which enables the gas to expand there and CR will tend to occupy preferentially more peripheral regions.

\subsection{CR streaming velocity}
\label{sec:CRstreaming}

Cosmic ray streaming is an important ingredient of the present work,
as mentioned above. We discuss here briefly the range of possibilities
for the CR streaming speed. But before we enter the detailed discussion, we summarise our main insights:
 
For a very turbulent medium, CR may be well trapped by magnetic irregularities produced by the turbulence. 
For a less turbulent high-$\beta$ plasma, the sound speed seems to be a reference speed.%
\footnote{In a low-$\beta$ plasma, the CR streaming velocity is linked to the Alfv\'en velocity, which exceeds the sound speed there. 
However, this can obviously not be true in a high-$\beta$ plasma. 
It would imply that in the limit of vanishing magnetic field strength the CR get completely immobile due to the vanishing Alfv\'en speed. 
Already intuition tells us that this must be wrong, since for disappearing magnetic fields, the coupling of CRs to the plasma gets weaker and therefore the CRs should stream faster.
Thus, there must be a characteristic velocity, below which the Alfv\'en velocity is not limiting the streaming velocity any more. 
Plasma physical arguments indicate that this is roughly the sound speed.}
However, in case
the turbulence and CR energy densities are low, even higher transport
speeds are possible.  Since we only know the streaming speed approximately, we will
parametrise it in terms of the sound speed.  Note, that complex magnetic topologies reduce the effective, macroscopic speed, and this effect 
should be stronger for turbulent clusters than for relaxed ones.

Now we investigate the different factors determining the streaming speed.
In doing so we shall clearly distinguish between two separate regimes,
one for merging clusters and the other for relaxed clusters.
In the first case the ICM is highly turbulent.  The turbulence is
transonic, i.e. the sonic Mach number is $M = \alpha_\rmn{tu} \ge 1$ and, for a typical
magnetic field strength of a few $\mu$G and sound speed of 10$^3$km/s,
highly super-alfv\'enic, i.e. the Alfv\'enic Mach number is $M_\rmn{A} =\vel_\rmn{tu}/\vel_\rmn{A} \sim
10$.
A small-scale dynamo is also operating at this stage. We assume that this has amplified 
the field to the observed level of a few microgauss from perhaps weaker fields.
Under these conditions the magnetic field lines are easily bent
by the inertia of the ICM motions, at least on scales where the
turbulence remains super-alfv\'enic. If the turbulence is injected on
scales $L_\rmn{tu} \simeq {\rm a~few} \times 10^2$ kpc, and follows
Kolmogorov's scaling, then super-alfv\'enic conditions persist down to
scales $L_\rmn{A} \simeq L_\rmn{tu} M_A^{-3}\simeq {\rm a~few} \times 10^2$ pc.
Under these conditions the CR particles are efficiently trapped in
between large amplitude magnetic field fluctuations by adiabatic
mirroring~\citep[e.g.][]{2005ppfa.book.....K}. So, at least
macroscopically, the CR particles effectively follow the advective 
motion of magnetic field lines.

As the merger process approaches dynamical equilibrium, the amount of
ICM turbulence drops considerably in the core regions, with typical
values of the gas speed $\vel_\rmn{tu}\sim {\rm a~few} \times 10^2$
km/s~\citep[see, e.g.][]{2005MNRAS.358..139F,2010arXiv1001.1170P}.
This changes the
dynamical picture in the core regions at a qualitative level.  In
fact, the turbulent motions are now subsonic and, more importantly,
trans-alfv\'enic, i.e. $M_A\gtrsim 1$. Under these conditions adiabatic
mirroring will be modest and the most effective way to limit 
the propagation speed of the particles is through the self-generated
turbulence.

In fact, CRs in the Solar neighborhood
are observed to have a very small anisotropy
(about 1 part in $10^4$). This is explained by
pitch-angle scattering of CRs by plasma waves, in
particular Alfv\'en waves, in our interstellar medium.
These waves can be generated
by the CRs themselves, by a streaming instability,
as they gyrate around and stream along the
magnetic field at a velocity faster than the 
Alfv\'en speed~\citep{1968ApJ...152..987W,1969ApJ...156..445K, 1971ApL.....8..189K}.
The waves traveling in the direction of the streaming CRs (and satisfying
a resonance condition for low frequency waves, 
$k_\rmn{wave}\, \gamma_\rmn{cr} \, \vel_\rmn{cr} \, \mu_\rmn{cr} = \Omega_\rmn{cr}$) are amplified by
this streaming instability. Here $k_\rmn{wave}$ is the wavenumber of the amplified
wave, $\vel_\rmn{cr}$ is the CR velocity, $\gamma_\rmn{cr}$ its Lorentz factor, $\Omega_\rmn{cr}$ its gyro frequency,
and $\mu_\rmn{cr} = \cos\theta_\rmn{cr}$, with
$\theta_\rmn{cr}$ the pitch angle, the angle between the CR velocity
and the magnetic field direction. For $\mu_\rmn{cr} \sim 1$, the amplified
waves have a wavelength of order the CR gyro-radius.
These self-generated waves can then resonantly scatter the CRs
and lead to an isotropisation of the pitch angle distribution in the
wave frame. Hence in the low plasma-$\beta$ interstellar medium
the CRs are expected to stream
with respect to the gas with the wave (or
Alfv\'en) velocity~\citep[see, e.g.,][for a detailed discussion]{2005ppfa.book.....K}.
The scattering also leads to a
transfer of momentum and energy from CRs to the waves,
which is in turn transmitted to the thermal gas, when
the waves damp out,  although this is a higher order effect than the 
pitch-angle scattering process and therefore slower.

A potential problem with the above picture is due to the fact
that in order to isotropise
their pitch angle distribution some CR particles need to
be scattered by the waves to have negative $\mu_\rmn{cr}$, or reverse
their parallel component of the velocity.
This can only happen if they get scattered through $\mu_\rmn{cr}=0$.
But for $\mu_\rmn{cr} \sim 0$, the resonance condition implies that
resonant waves will have a large $k_\rmn{wave}$. In a high-$\beta$ plasma,
such as the ICM, such large $k_\rmn{wave}$ waves can suffer strong 
resonant damping by thermal protons~\citep{1979ApJ...228..576H}. Thus to maintain them at a
significant level one needs a larger streaming speed.
~\cite{1979ApJ...228..576H} then deduced that, in a high-$\beta$ plasma,
CRs can then stream at a speed of order or greater than the
ion sound speed, which in this case will be larger than the
Alfv\'en speed.

Possible processes can help in crossing the $\mu_\rmn{cr}=0$ gap, include
mirroring on
long wavelength waves~\citep{1981A&A....98..161A,2001ApJ...553..198F}, and resonance
broadening which relaxes the resonant
condition~\citep{1981A&A....98..161A,2008ApJ...673..942Y}.  As for the
former case, long wavelength waves -- responsible for the mirroring -- are
assumed to be generated by the streaming CR particles themselves,
albeit at a smaller pitch angle.
However, they can also be produced as a result of residual turbulent
motions which we know exist even in the ICM of relaxed
clusters~\citep{2005MNRAS.358..139F,2010arXiv1001.1170P}.  In any
case~\cite{2001ApJ...553..198F} find that the contribution from
mirroring to the diffusion coefficient is important but not dominant.
The streaming velocity, $\vel_\rmn{str}$, can then be found by balancing (A)
the growth rates of the resonant waves produced by the streaming
particles and the (nonlinear) damping rates, which is proportional to
the amount of resonant waves; and (B) by noting that the streaming
velocity is determined by the anisotropic component of the CRs which
is inversely proportional to the amount of resonant waves responsible
for pitch angle scattering. This is basically the argument followed
by~\cite{2001ApJ...553..198F}. If we apply this argument to the case
to the ICM we find that, among others, the streaming velocity scales
inversely with the square-root of the number density of the CR
themselves.  If the CR energy is at the level of ten percent or so,
then the streaming velocity is of order of the thermal speed of the
ions, i.e. we find that Eq.~\ref{vst::eq} is approximately valid.
If the CR population is substantially lower than that, e.g.  as
in the case of CRe or for CRp after  a result of their
streaming motions, the streaming velocity can become significantly
larger than the sound speed.

We should note that during both the turbulent and quiescent 
phases, the magnetic field is not static. When the gas is turbulent,
there is a dynamo, which not only amplifies
the field but also makes it spatially and temporally intermittent.
A place which has strong field at some time will, after eddy turn
over time, have weaker field and also possibly with different connectivity.
However this does not affect the streaming instability drastically 
as the growth and damping rates of the waves will generally occur 
at a much faster rate than the growth rate of the small-scale dynamo.
But it will affect the long time behaviour of CRs, because of 
the constantly changing and complex connectivity of the field, although the 
consequences of such changes for CR propagation have yet to be studied.
More interesting is what would happen when there is no longer
merger-driven turbulence. The field itself then generates
decaying MHD turbulence, with the kinetic and magnetic energy 
decaying as a power law in time, and the coherence length of the field
increasing also as a power law \citep{2006MNRAS.366.1437S}.
Such an increased correlation length and simpler connectivity could
make it easier for the CRs to propagate throughout the cluster,
albeit limited by the streaming velocity.

\subsection{Interplay of advection and streaming}
\label{sec:closedBstruct}
\begin{figure*}
 \includegraphics[bb=0 0 653 181,width=\textwidth]{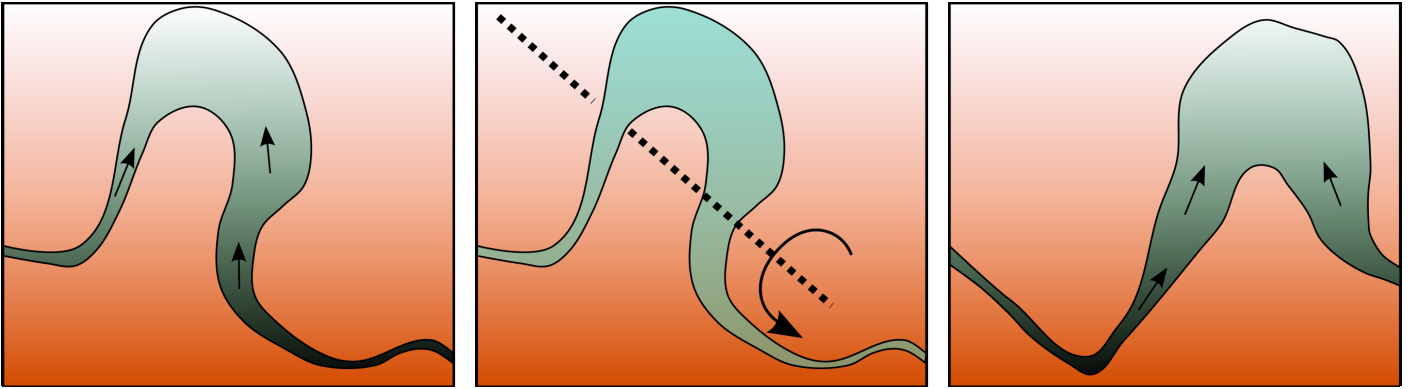}
 % sketch1.eps: 1179648x0 pixel, 300dpi, 9987.69x0.00 cm, bb=0 0 653 181
 \caption{Sketch of the interplay of CR streaming and turbulent advection for a single flux tube in a stratified atmosphere with gravity pointing downwards. \textbf{Left:} The dense CR at the center stream along the tube towards the CR depleted regions at larger atmospheric height. \textbf{Middle:} CR streaming stops as soon as a homogeneous CR space density is achieved. A turbulent eddy (represented by its angular momentum axis) starts to turn the magnetic structure upside down. \textbf{Right:} The former outer parts of the flux tubes are compressed at the center, and harbor now an overdense CR population, whereas the former inner parts are expanded at larger atmospheric scale height and therefore have now an underdense CR population. Again CR streaming sets in.}
 \label{fig:sketch1}
\end{figure*}

First, we consider an isolated magnetic flux tube with CRs confined to it to illustrate the interplay of advection and streaming with a basic picture. This will be generalised later on in Sect.~\ref{sec:expectedCRprofile} to a more complex situation including CRs escaping from magnetic structures. Focusing on the CR transport along a single flux tube embedded in the ICM, we start with the idealised picture of a static flux tube frozen into a static plasma as shown in Fig. \ref{fig:sketch1} on the left. Any central concentration of CR will escape due to CR streaming on a timescale of
\begin{equation}
 \tau_\mathrm{st} = \frac{L_B}{\vel_\mathrm{st}} = \frac{\chi_B}{\alpha_\mathrm{st}}\, \frac{L_\mathrm{tu}}{c_\mathrm{s}}.
\end{equation}
This leads to a homogeneous CR distribution within the flux tube (Fig. \ref{fig:sketch1}, middle). Turbulence turns the magnetic structure upside down on half an eddy turnover time $\tau_\mathrm{tu} = {\pi\, L_\mathrm{tu}}/(2\, \vel_\mathrm{tu})$. This is comparable to, or less than, the CR escape time, 
\begin{equation}
 \frac{\tau_\mathrm{st}}{\tau_\mathrm{tu}} 
\equiv \gamma_\rmn{tu}
\sim \chi_B  \frac{\alpha_\mathrm{tu}}{\alpha_\mathrm{st}} \sim \mathcal{O}(1),
\end{equation}
and 
thus a good fraction of the CR from larger radii will be compressed  towards the center, from where they again start streaming to larger radii. 

The transonic turbulence is therefore able to maintain a centrally enhanced CR density by pumping expanded CR populations downwards. As soon as the turbulent velocities become significantly subsonic, this pumping becomes inefficient, since the streaming will be faster than the advection. At this point a nearly constant volume density of CR establishes within a closed flux tube, meaning that most CR are residing at larger cluster radii. However, the CR density can again become centrally enhanced, if the turbulence becomes strong again, e.g. during the next cluster merger. 

During the phase of transonic turbulence, we regard two Lagrangian volume elements which are exchanged radially by an eddy, one starting at small radius with volume $V_1$ and the other at large radius with volume $V_2 = X\,V_1$, with $X>1$. We assume a relaxed CR population with power law spectrum as in Eq. \ref{eq:CRspec} being present in them, with $C_1  = C_2 = C$. Now, $V_2 \rightarrow V_2' = V_1$, the enclosed CR get compressed and gain momentum  adiabatically according to
\begin{equation}
 p_2 \rightarrow p_2' = X^{\frac{1}{3}} \, p_2. 
\end{equation}
The compressed CR population is again described by  Eq. \ref{eq:CRspec}, but now with
\begin{equation}
 C_2' = C\,X^{\frac{\alpha +2}{3}}.
\end{equation}
Correspondingly, the expanded CR population in $V_1'$ has a power-law normalisation constant of $C_1' = C\,X^{-({\alpha +2})/{3}}$. A simple swap of the two volumes can generate a CR density contrast (at fixed momentum) of $C_2' /C_1' = X^{({2\,\alpha +4})/{3}} \approx 2^3 = 8$ from a previously homogeneous distribution if the ratio of the initial volumes was only $X=2$. 

The complete process of advective compression and expansion via CR streaming is, from the perspective of the CRs, completely adiabatic. The energy losses during the streaming expansion phase will be exactly compensated during the advective compression phase. This might surprise, since the dissipation of the excited waves is certainly not a reversible process (for any practical purpose). But the energy injected into the wave fields comes from adiabatic work done by the expanding CR gas. All what the CRs \textit{feel} are the expanding \textit{walls} of the self-generated plasma waves, and the converging magnetic fields confining them when they are dragged back inwards by a downwards directed flow.

However, from the perspective of the turbulence, the process is dissipative. 
Not all of the kinetic energy invested into CR compression will be returned to the kinetic energy budget during a gas expansion phase, but some fraction is transferred via plasma wave generation and damping into thermal heat. 
Thereby this process contributes to the damping of turbulent flows as well as to the heating of the ICM. 
Since the CR can release their energy at different locations than where they got it via adiabatic compression in the turbulent flow, the spatial distribution of their heating power will in general differ from that of the turbulence. 
Since CR actively propagate preferentially to regions of low CR densities, their heating profile can be expected to be more regular than that of the probably intermittent turbulent cascades. 

\subsection{Diffusive acceleration}
\label{sec:acc}
CR diffusion will always be present at some level, although it is probably a less efficient transport mode compared to streaming. 
It has, however, some other advantage in that this propagation mode does not lead to adiabatic energy losses of the CR.
Diffusion also tends to establish a constant CR density per volume as streaming does, but with a positive energy balance if counteracted by turbulent advection. If the volume expansion factor $X$ can be split into a (dominant) streaming and (subdominant) diffusion part, we obtain
\begin{equation}
 X = X_\mathrm{str}\, X_\mathrm{diff}.
\end{equation}
If this is identical to the (inverse of the) subsequent advective compression $X_\mathrm{adv}$, then a net energetisation of the CR population happens. This is described by
\begin{equation}
 \frac{dC}{dt} \approx C\, \frac{X_\mathrm{diff}^{\frac{\alpha-1}{3}}}{\tau_\mathrm{tu}},
\end{equation}
and leads to an exponential energy gain of the CR. The individual particles gain per cycle the momentum
\begin{equation}
p \rightarrow p' = p \, X_\mathrm{diff}^{\frac{1}{3}}.
\end{equation}
In case some fraction $\eta_\mathrm{loss}$ of CR gets lost from the turbulent region per cycle, the spectrum evolves towards a power law distribution with spectral index
\begin{equation}
\alpha =   1 + {\eta_\mathrm{loss}}\,{X_\mathrm{diff}^{-\frac{1}{3}}},
\end{equation}
assuming a continuous injection of low energy CRs is provided. For a low loss fraction $\eta_\mathrm{loss} < X_\mathrm{diff}^{{1}/{3}}$, the asymptotic spectrum can even become harder than the canonical value $\alpha=2$ of diffusive shock acceleration in the strongest non-relativistic shocks (and in the test-particle limit).

Given the final duration of the transsonic ICM turbulence of $1\,\rmn{Gyr}\sim 10 \, \tau_\mathrm{eddy}$ (partly because the virializing dark matter flows continue to stir turbulence for the dynamical time of a cluster), and the limited pool of low energy CRs, due to the severe Coulomb losses in the subrelativistic regime, such a spectrum is probably not established in typical galaxy clusters. In a cool core, a slightly larger number of eddy turn overs might be realistic, however, the escape probability of the accelerated CRs into the wider ICM might be large, so that the accelerated spectrum might be steep. 

This discussion should have made clear that some CR energetisation can be expected if CR diffusion contributes substantially to the active CR propagation. 
In order for this to happen, the poorly known CR diffusion coefficient has to be on the right order of magnitude to have the diffusion speed being a non-negligible fraction of the advection speed. 
Different to the case of CR streaming, we are not aware of any natural explanation for this. 
In case of CR streaming in post-merger clusters, the cluster turbulence velocity as well as the CR streaming speed are each linked to the sound speed and therefore comparable.

\subsection{Expected CR profiles}
\label{sec:expectedCRprofile}

We now generalise the picture to full turbulent advection, complex magnetic topology, and CR transport. 
What are the expected CR profiles in galaxy clusters that are established by the interplay of advection and propagation of CR? Is the process limited to operate only when the assumption of strict CR confinement to flux tubes is valid?

In order to answer these questions, we assume CR diffusion to enable the CR to change between magnetic flux tubes and thereby find paths to more peripheral regions. Thus the CR propagation needs not to stop after $L_B$ but can reach even the outskirts of galaxy clusters, where the infall of matter onto the cluster behind the accretion shocks prevents further escape.%
\footnote{This infall has typical Mach numbers of a few to ten. 
Thus CR would need to stream with highly supersonic velocity against this matter stream in order to escape, which does not look very likely. 
However, even if some CR would escape this way, we do not believe that the effect would be sufficiently large to be necessarily included in our rough modeling. 
First, as we will see in Sect. \ref{sec:timescales}, the periods of low cluster turbulence seem to be sufficiently long that the spatial CR distribution can relax significantly, but not fully. 
Thus, the bulk of the CR distribution never propagates to the cluster boundary. 
Second, in case high energy cosmic rays manage to escape against the cluster accretion flow, many more lower energy CR 
would manage to cross the accretion shock, but would be swept back, thereby be compressed and further energetised at the shock wave. 
This implies, that the cluster boundary is more a CR energy source, than a CR leak.
In order to keep our discussion simple, we ignore both, potential CR escape through and CR injection and re-acceleration at the cluster accretion shock wave.}

Without turbulent advective transport counteracting, we would expect a completely homogeneous CR distribution within the galaxy cluster to establish itself after some time. 
Since the required perpendicular CR diffusion steps might be slow, the macroscopic 
$\bar{\alpha}_\mathrm{st}$ should be smaller than the microscopic one $\alpha_\mathrm{st}$ of the previous sections. If we model the diffusive transport through bottlenecks as a spatial decrement in $\alpha_\mathrm{st}(r)$, where $r$ is the (radial) coordinate parallel to the CR gradient, we find 
\begin{equation}
 \bar{\alpha}_\mathrm{st} = \left(\frac{1}{L} \int_0^{L\,\gg L_B}\!\!\! \!\!\! dr \frac{1}{\alpha_\mathrm{st}(r)} \right)^{-1}.
\end{equation}
Thus, a few bottlenecks with $\alpha_\mathrm{st}(r)\ll 1$, at which the CRs spend most of their time with low transport velocity, can reduce the macroscopic speed by a significant factor to be well below the sound speed. In the following we simply write $\alpha_\rmn{st}$ instead of $\bar{\alpha}_\mathrm{st}$ for the macroscopic streaming Mach number.

If the diffusion of the CRs out of magnetic bottlenecks dominates the transport time, any energy dependence of the effective diffusion coefficient imprints itself on the macroscopic streaming speed. 
We are not going to model this in the following, however, we note, that the energy dependence of the transport can lead to spatially varying spectral indices of a CR population. 

The cluster-wide CR profile relaxation time of $\tau_\mathrm{relax} \sim R_\rmn{as}/\vel_\rmn{st}\sim 10 \, r_\rmn{c}/\vel_\mathrm{st} \sim 2 \, \mathrm{Gyr}/{\alpha}_\mathrm{st}$ can be comparable to the cluster age (where $R_\rmn{as} \approx 10\, r_\rmn{c}$ is the radius of the accretion shock confining the CR population within the cluster). Thus, turbulent CR advection could be very essential to maintain a centrally peaked CRp profile, if we want to explain radio halos with the hadronic model, or a centrally peaked low-energy CRe population, for re-acceleration models to be operative.
CR propagation aims at establishing a spatially homogeneous CR distribution, but what is the preferred profile of adiabatic advection? 

To work this out, we use the following assumptions.
\begin{enumerate}
 \item The cluster is characterised by  a mean pressure profile. For the sake of simplicity, we ignore pressure fluctuations due to the turbulence.
\item CR propagation operates on small scales, permitting CR exchange between nearby gas volume elements, but not on large scales so that the profile is still determined by the advective transport.
\item Magnetic bottlenecks reduce the macroscopic CR streaming speed to a level that it is irrelevant for the global profile. This assumption will be alleviated later on.
\end{enumerate}
Under these conditions turbulent advection will completely dominate the CR profile. 

Whenever two volume elements come close at some radius, they might exchange CR, tending to establish a constant CR population in any given radial shell.\footnote{Or more precisely, on any surface of constant pressure, which we identify from now on with radial spheres for simplicity.} During radial advective transport from radius $r$ to $r'$ the ICM gas with the enclosed CR  is compressed or expanded by a factor $X(r \rightarrow r') = (P(r')/P(r))^{1/\gamma}$, where $P(r)$ is the pressure profile and $\gamma=5/3$. 
The CR rest-mass density $\varrho(\vec{r})= m\,\int dp\, f(\vec{r},p)$  (or any other advected quantity)
therefore wants to establish -- under the influence of advection alone -- a profile according to 
\begin{eqnarray}
\label{eq:varrhowithP}
 \varrho(r) &=& \varrho_0\, \left( \frac{P(r)}{P_0} \right)^{\frac{1}{\gamma}}=   \varrho_0 \, \eta(r),
\end{eqnarray}
where $\eta(r) = (P(r)/P_0)^{1/\gamma}$ is the advective CR target profile.
This is the only steady state profile which fulfills the requirement $X(r \rightarrow r') = \varrho(r')/ \varrho(r)= (P(r')/P(r))^{1/\gamma}$ for any $r$ and $r'$.
The CR normalisation develops therefore a radial profile 
\begin{eqnarray}
\label{eq:C(r)withP}
 C(r) &=& C_0\, \left( \frac{P(r)}{P_0} \right)^{\frac{\beta_\mathrm{cr}}{\gamma}} 
= C_0\, \left(  \eta(r) \right)^{\beta_\mathrm{cr}}
\!\!, \;\; \mbox{with} \nonumber\\
 \beta_\mathrm{cr} &=& \frac{\alpha+2}{3} \approx 1.33\cdots 1.67 .
\end{eqnarray}
This profile is strongly peaked at the cluster center, and is very much in contrast to the spatially constant profile  with $\beta_\mathrm{cr}=0$ that CR propagation seeks to establish.

\begin{figure*}
 \includegraphics[bb=25 180 550 600,width=0.49\textwidth]{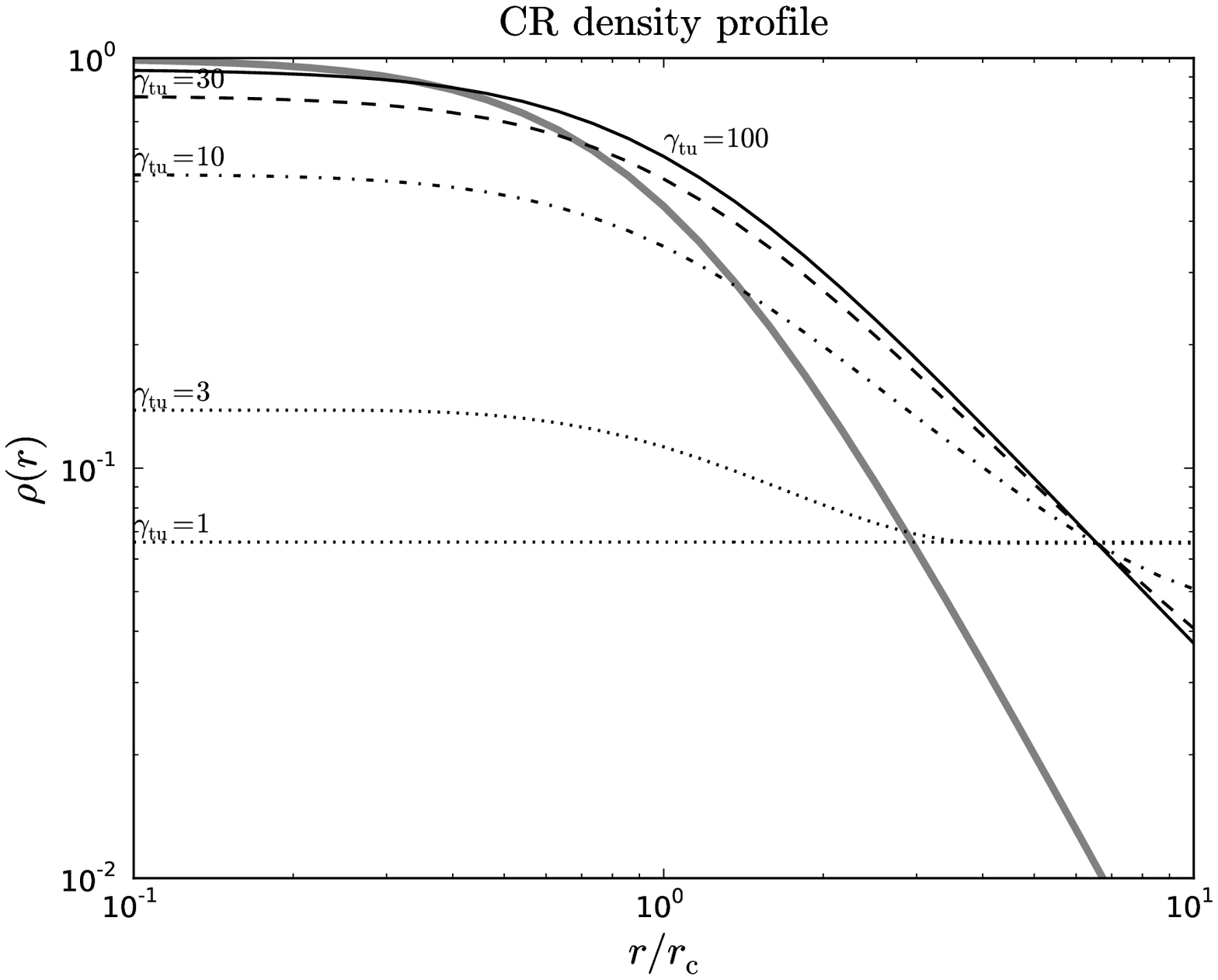}
 % density.eps: 0x0 pixel, 300dpi, 0.00x0.00 cm, bb=18 180 594 612
\includegraphics[bb=25 180 550 600,width=0.49\textwidth]{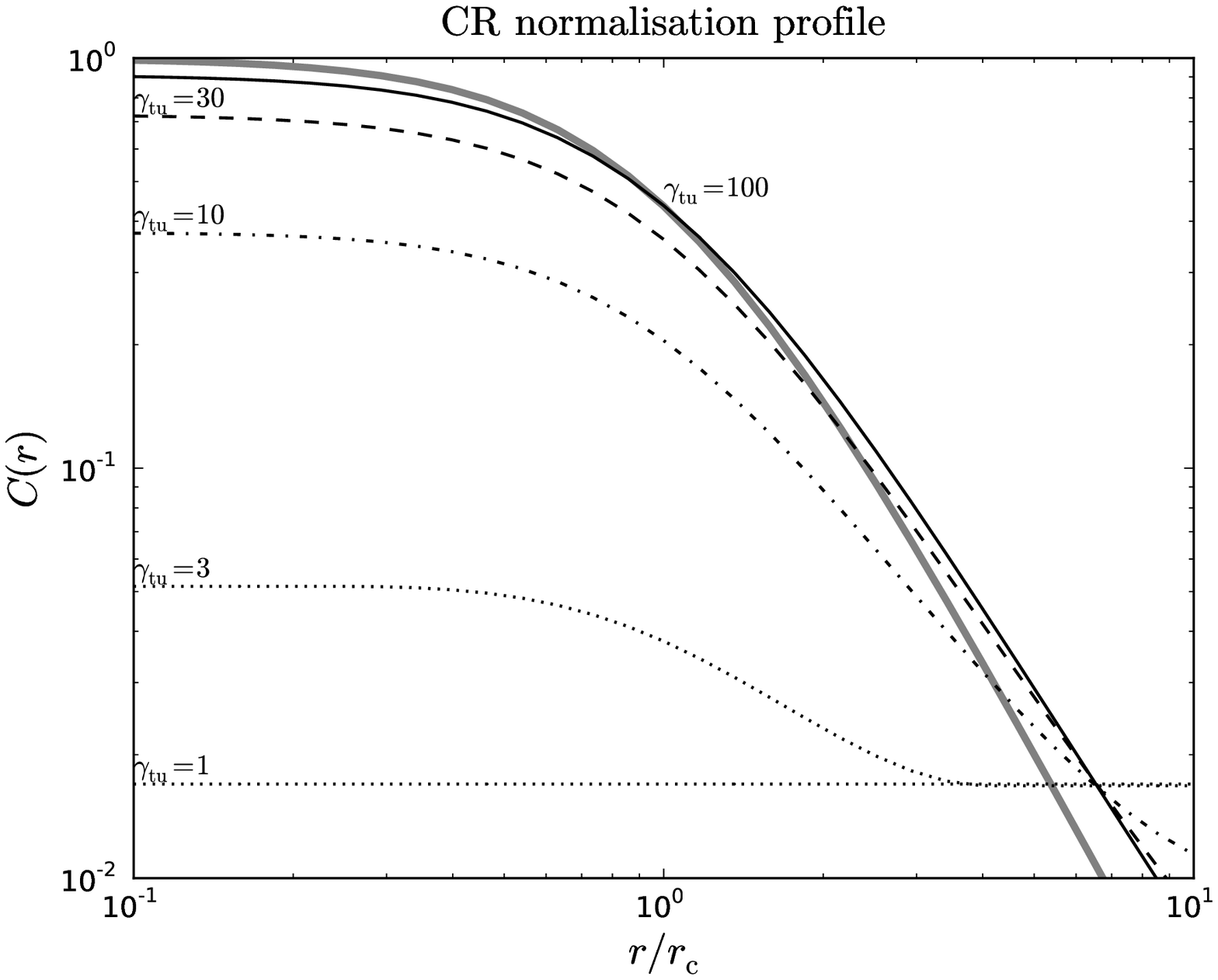}
 \caption{\textbf{Left:} CR density profiles for $\gamma_\rmn{tu} = 1, 3,\,10, \, 30,$ and $ 100 $ (from bottom to top at small radii) including the same number of CRs each. Profiles are normalised to $\varrho(0)|_{\gamma_\rmn{tu}=\infty}$. Also the more narrow gas density profile is shown (thick grey line). \textbf{Right:} CR normalisation profiles for the same parameters and the gas density profile.  ($\beta_\rmn{cl}=0.8$, $R_\rmn{as} =10$, $\alpha=2.5$).}
 \label{fig:density}
\end{figure*}

In reality, CR propagation and advection will both shape the CR profile. The ratio of their transport coefficients will determine the exact equilibrium shape. We will argue that during the transsonic turbulence of a cluster merger, we expect both processes to operate on comparable strength and therefore we expect a profile which is centrally peaked. A Gyr after a cluster merger, turbulence decays and CR streaming becomes more dominant, leading to a flattened profile implying a negligible CR population in the cluster center compared to the turbulent advection case. Thus, we expect a large variety of CR profiles being present, depending on the turbulent history of the cluster.  This must have implications on radiative signatures of CRs in clusters.

In the following, we want to get some analytical insight into the dependence of the CR profile on the turbulence level. All we aim for is a rough model, which captures the essential dependencies, and leaves a more accurate treatment to numerical simulations. The CR continuity equation for $\varrho$ in the absence of sources and sinks can be written as 
\begin{equation}
 \frac{\partial  \varrho}{\partial t} + \vec{\nabla}\cdot(\vvel\, \varrho) = 0,
\end{equation}
with $\vvel =  \vvel_\mathrm{ad} + \vvel_\mathrm{di} +  \vvel_\mathrm{st}$ the CR transport velocity, which is composed by advective ($\vvel_\mathrm{ad}$), diffusive ($\vvel_\mathrm{di}$), and steaming ($\vvel_\mathrm{st}$) transport velocities. 

We characterise the passive, advective  transport via turbulence as an additional diffusion process with diffusion coefficient
\begin{equation}
 \kappa_\mathrm{tu} = \frac{L_\mathrm{tu}\, \vel_\mathrm{tu}}{3}.
\end{equation}
In a stratified atmosphere the effective and averaged advective velocity of this (passive, macroscopic) diffusion is given by
\begin{equation}
 \vvel_\mathrm{ad} =   - \kappa_\mathrm{tu} \, \frac{\eta}{\varrho} \, \vec{\nabla} \frac{\varrho}{\eta} =  -\kappa_\mathrm{tu} \, \vec{\nabla} \ln \left( \frac{\varrho}{\eta}\right) .
\end{equation}
What is a bit unusually here is the appearance of the target density profile $\eta(\vec{r}) = (P(\vec{r})/P_0)^{1/\gamma}$, which ensures that the advective diffusion alone tries to establish the steady state CR density profile we derived above in Eq. \ref{eq:varrhowithP}. Its appearance can be understood as follows:

If turbulent advection is dominating the solution, we should find $\varrho/\varrho_0 = \eta$ in the equilibrium configuration, according to Eq. \ref{eq:varrhowithP}. Thus the effective gradient in the CR population, which drives the turbulent transport, has to vanish. This can only be achieved by a term proportional to $\vec{\nabla} (\varrho/{\eta})$. Already by dimensional reasoning one sees that the prefactor has to be $- \kappa_\mathrm{tu} \, {\eta}/{\varrho}$.

The counteracting streaming and diffusive CR propagation can also be described by velocities. The streaming velocity has magnitude $\vel_\rmn{st}$ and is anti-parallel to the CR space density gradient. Active CR diffusion can be described very similar to turbulent diffusion, just with a constant target space density. Thus we have
\begin{eqnarray}
 \vvel_\mathrm{st} &=&  - \vel_\mathrm{st}\, \frac{\vec{\nabla} \, \varrho}{|\vec{\nabla} \, \varrho|},\;\;\mbox{and}\\
\vvel_\mathrm{di} &=&   - {\kappa}_\mathrm{di}\, \, \frac{1}{\varrho} \, \vec{\nabla} \varrho =  - \kappa_\mathrm{di}\, \vec{\nabla} \ln(\varrho),
\end{eqnarray}
with ${\kappa}_\mathrm{di}$ the macroscopically averaged CR diffusion coefficient, which depends on $\kappa_\|$, $\kappa_\perp$ and the magnetic topology in a nontrivial way. We are not trying to give a concrete value for  ${\kappa}_\mathrm{di}$ from first principles, however, we assume that diffusive transport is subdominant but non-zero, since streaming over distances $> L_B$ requires that CR switch their guiding magnetic field lines. If diffusive CR transport would be significant, it would correspond to an anisotropy in the CR phase-space function, which would amplify plasma waves on which the CR would scatter. Shortly, the diffusive transport is limited by the streaming regime.

The CR space density becomes stationary for $\vvel = \mathbf{0}$, and this reads in spherical symmetry with radially outstreaming CRs
\begin{equation}
\label{eq:transpSpherical}
  {\vel}_\mathrm{st} =  \kappa_\mathrm{tu} \, \frac{\partial}{\partial r} \ln \left( \frac{\varrho}{\eta}\right)  + \kappa_\mathrm{di}\,  \frac{\partial}{\partial r}  \ln(\varrho) .
\end{equation} 
This is solved by
\begin{eqnarray}\label{eq:CRprofileSol}
 \varrho(r) &=& \varrho_0 \, \exp \left( \int_0^{r} \!\! dr'\, \frac{\vel_\mathrm{st}+ \kappa_\mathrm{tu} \,  \frac{\partial}{\partial r'} \, \ln \eta}{ \kappa_\mathrm{tu}  + \kappa_\mathrm{di} }\right)\nonumber\\
&= &\varrho_0 \, \eta(r)^{\beta_\varrho} \, \exp \left( {\frac{r}{r_*}} \right),\;\; \mbox{with} \\
\beta_\varrho &=& \frac{\kappa_\mathrm{tu}}{ \kappa_\mathrm{tu}  + \kappa_\mathrm{di} }
= \frac{1}{ 1  + \kappa_\mathrm{di}/\kappa_\mathrm{tu} } <1
 ,\;\; \mbox{and} \nonumber\\
r_* &=&  \frac{ \kappa_\mathrm{tu}  + \kappa_\mathrm{di} }{\vel_\mathrm{st}} = \frac{\kappa_\mathrm{tu} }{\beta_\varrho \,\vel_\mathrm{st}} =  \frac{\alpha_\mathrm{tu}\,\chi_\mathrm{tu} }{3\, \beta_\varrho \,{\alpha}_\mathrm{st}}\, r_\rmn{c}.\nonumber
\end{eqnarray}
In the second line $\vel_\mathrm{st}$, $\kappa_\mathrm{tu}$, and $\kappa_\mathrm{di}$ were assumed to be spatially constant. 
In the following, we adopt a standard cluster beta-profile for the thermal electron density $n(r)$ 
\begin{equation}
 n(r) = n_0 \, \left( 1 + \frac{r^2}{r_\rmn{c}^2}\right)^{-\frac{3}{2}\, \beta_\mathrm{cl}}.
\end{equation}
Here we assume for simplicity a typical value for $\beta_\mathrm{cl} \sim 0.8$ and a constant temperature profile.\footnote{We note that this can be easily generalised to a more realistic declining temperature profile in the peripheral cluster regions and emphasize that our results do not depend on the exact value of the outer slope of the density profile and the external cluster regions beyond $R_{500}$ -- at least for streaming distances that are small compared to the virial radius.} With these simplifying assumptions $P(r)/P_0 = n(r)/n_0$ and 
\begin{equation}
 \eta(r)= \left( 1 + \frac{r^2}{r_\rmn{c}^2}\right)^{-\frac{3\, \beta_\mathrm{cl}}{2\, \gamma}}.
\end{equation}
Thus the CR density profile is a bit flatter than the thermal density profile. 

The solution of Eq. \ref{eq:CRprofileSol} is only physical for this profile between 
\begin{equation}\label{eq:r+-}
 r_\pm = \frac{3\,\beta_\mathrm{cl}\, \beta_\varrho }{2\,\gamma}\, r_* \, \left( 1 \pm \sqrt{1- 
\left( \frac{2\,r_\rmn{c}\,\gamma}{3\,\beta_\mathrm{cl}\, \beta_\varrho\,r_*}\right)^2}\right) \,,
\end{equation}
since at these radii $\vvel_\mathrm{st}$ changes sign, which we did not model in Eq. \ref{eq:transpSpherical}. The CR profile outside $r_- < r < r_+$ is actually non-stationary. Here, we just set $\varrho(r) = \varrho(r_\pm) $ for $r<r_- $ and $r>r_+$, respectively, since the dominating streaming in these regimes wants to establish a constant CR volume density, which we take as an acceptable approximation to the time-averaged density there. These radii are approximately given, in case $r_- \ll r_+$, by
\begin{eqnarray}
 r_- &\approx&  \frac{r_\rmn{c}^2\gamma}{3\,\beta_\mathrm{cl}\, \beta_\varrho\,r_*} =  \frac{{\alpha}_\mathrm{st}\,\gamma}{\alpha_\mathrm{tu}\,\chi_\mathrm{tu} }\, \frac{r_\rmn{c} }{\beta_\mathrm{cl}},\;\; \mbox{and} \nonumber\\
r_+ &\approx& \frac{3\, \beta_\mathrm{cl}\, \beta_\varrho}{\gamma} \, r_* =  \frac{\alpha_\mathrm{tu}\,\chi_\mathrm{tu}\,  \beta_\mathrm{cl}}{{\alpha}_\mathrm{st}\,\gamma}\, r_\rmn{c}\approx \frac{r_\rmn{c}^2}{r_-}.
\end{eqnarray}
A centrally enhanced CR profile exists only if $r_\pm$ are real (see Eq. \ref{eq:r+-}), which translates into the necessary condition 
\begin{equation}
 \gamma_\mathrm{tu} = \frac{\alpha_\mathrm{tu}\,\chi_\mathrm{tu}}{3\,{\alpha}_\mathrm{st}} = \frac{\kappa_\rmn{tu}}{\vel_\rmn{st}\,r_\rmn{c}} = \frac{{\tau}_\rmn{st}}{\tau_\rmn{ad}}> \frac{2\,\gamma}{3\,\beta_\mathrm{cl}}\approx 1.7
\end{equation}
is fulfilled. It means that in order to develop a centrally enhanced CR profile, turbulence has to be above some critical value (comparable to the  sound speed), and the CR streaming must be inhibited by magnetic topology constrains to a value significantly below the sound speed. 
More specifically, the timescale $\tau_\rmn{ad}= r_\rmn{c}^2/\kappa_\rmn{tu} =3\,\tau_\rmn{tu}$  for turbulent diffusion over the distance $r_\rmn{c}$ should be significantly shorter than the macroscopic streaming timescale ${\tau}_\rmn{st}= r_\rmn{c}/{\vel}_\rmn{st}$ for the same distance. 
This implies that magnetic bottlenecks are critical in lowering the microscopic streaming velocity of CR by some finite factor. In order that the macroscopic streaming does not disappear (and we do not have a switch off mechanism for halos) some cross-field diffusion must be present. 
Since we expect then also some diffusion along the magnetic field lines, we can expect some level of diffusive acceleration of CRs as discussed in Sect. \ref{sec:acc}.

However, in order to also have a stationary situation for the CR spectra, we assume that re-acceleration is negligible, i.e. that $\kappa_\mathrm{di}\ll\kappa_\mathrm{tu}$ and therefore $\beta_\varrho \approx 1$. Since then both dominating transport mechanisms, turbulent advection and CR streaming, are adiabatic in nature,  we can translate the CR density profile into a profile of the spectral normalisation simply by using $C(r) = C_0 \, (\varrho(r)/\varrho_0)^{\beta_\rmn{cr}}$. 

For our beta-profile cluster this yields
\begin{eqnarray}
 C(r) &=&  C_0 \, \left( 1 + \frac{r^2}{r_\rmn{c}^2} \right)^{- \beta_\mathrm{C}}\,\exp\left(\frac{r}{r_*}\, \beta_\rmn{cr}\right)
, \;\; \mbox{with} \\
\beta_\mathrm{C} &=& \frac{3}{2}\, \beta_\mathrm{cl} \, \beta_\varrho\, \beta_\rmn{cr},\;\mbox{and}\;\; \beta_\rmn{cr} = \frac{\alpha+2}{3}\nonumber
\end{eqnarray}
within $r_-< r< r_+$ and 
\begin{equation}
 C(r) = C(r_\pm) 
\end{equation}
for $r>r_+$ and $r<r_- $, respectively. 

When we compare the different stages of the same galaxy cluster during its evolution during and after a cluster merger, 
we normalise the CR populations to have a constant total CR number
\begin{equation}
 N_\rmn{CR} = 4\,\pi\,\int_0^{R_\rmn{as}} dr\,r^2\,\frac{\varrho(r)}{m}.
\end{equation}
This is a conservative assumption, since during a cluster merger additional CRs will be injected, so that the CR content of pre-and post merger cluster should differ by some margin. The resulting CR density profiles are displayed in Fig. \ref{fig:density}.

We will use this toy model in the following to illustrate the effect of the CR transport processes on CR signatures in clusters of galaxies. It is certainly a very idealised picture in many respects. For example, the quasi stationary configuration implied is probably never reached in real clusters. The finite lifetimes of any cluster weather phase probably leads to a constantly changing CR profile. Nevertheless, it can be regarded as an extreme model, being the opposite extreme to the case of strict CR confinement in finite magnetic structures discussed in the previous section. 
Thus we have two scenarios, which probably bracket the physical reality in galaxy clusters:
\begin{enumerate}
 \item CR are confined in closed magnetic structures, which due to their radial extension, connect regions with some pressure ratio and therefore can be characterised by a CR compression ratio $X = (P_1/P_2)^\frac{1}{\gamma}$, as discussed in Sect. \ref{sec:closedBstruct}. We refer to this as the case of \textbf{confined CR}.
\item CR are able to stream relative freely (with reduced effective streaming velocity) throughout the whole cluster volume, and manage to establish the stationary profiles described in this Sect. \ref{sec:expectedCRprofile}. We refer to this as the case of \textbf{mobile CR}.
\end{enumerate}
In the following, we discuss the observational signatures of these two extreme cases. But first, we should look at the time-scales required to establish those steady-state profiles.

\subsection{Time scales}
\label{sec:timescales}

The profiles calculated in the previous section are mere illustrations of how the CR distribution might look like. 
In reality, the CR profiles will always be in an intermediate state between such profiles, since the cluster turbulence is never stationary and also does not exhibit the same velocities at all radii. 
In order to see if these profiles can be used to guide our intuition, we have to see if the real CR distribution is able to change significantly within a sufficiently short time.

\textbf{Radio halo switch off:} We assume that at a certain moment CR streaming has become the dominant transport mechanism, removing CRs especially from the center. In order to diminish or extinguish a radio halo, the central CR density in a cluster might fall by a factor of $X=2$ or $X=10$, respectively. If we approximate the central CR distribution as a homogeneous sphere with radius $r_\rmn{c}$, this requires an increase of the radius to $r = X^{1/3}\,r_\rmn{c} = 1.26\,r_\rmn{c}$ or $2.15\, r_\rmn{c}$, respectively. Thus, the CR have to stream a distance of $0.26\ldots1.15\, r_\rmn{c}$, which takes them $t = (0.06\ldots 0.28)/\alpha_\rmn{st}$ Gyr, respectively, for a sound speed of $c_\rmn{s}=1000\,\rmn{km\, s^{-1}}$ and $r_\rmn{c}=250\,$kpc. Note, that the CR normalisation drops by a larger factor $X^{(\alpha +2)/3} \approx X^{1.5} = 3 \ldots 30$, respectively, due to the effect of adiabatic energy losses.

As we will see in Sect. \ref{sec:radioHaloes}, this corresponds to a decrease in radio flux in the hadronic model by at least a factor  $X^{1.5\ldots 3} = 3\ldots 1000$, depending on the central magnetic field strength and the expansion factor. Thus, a switch off of cluster radio halos on timescales below a Gyr can be expected even in case that $\alpha_\rmn{st}<1$, as long it is not too low ($\alpha_\rmn{st}> 0.06\ldots 0.26$). 

In the re-acceleration model, the radio halo switches off with the radio halo CRe cooling timescale of about 0.1 Gyr the moment the re-acceleration stops to operate, irrespective if there is CRe transport from the outside or not.

\textbf{Radio halo switch on:} Both radio halo models, the hadronic and the re-acceleration one, require a seed CR population to form a radio halo, either protons or electrons, respectively. 
Again, transport over a core radius is required in order to pump CRs into the cluster center. 
The timescale for this is $t = r_\rmn{c}^2/\kappa_\rmn{tu}= 0.73/(\alpha_\rmn{tu}\,\chi_\rmn{tu})$ Gyr, and therefore sufficiently short if the turbulence is sufficiently transsonic. 
The initial transport of CR into the cluster center during a merger is probably poorly described by our turbulent diffusion model. 
It might be proper to assume that the CR just have to be advected by a coherent flow through the cluster core, so that $t = 2\,r_\rmn{c}/\vel_\rmn{tu} = 0.5/\alpha_\rmn{tu}\,$Gyr. 
However, realistic descriptions should use 3-d numerical hydrodynamical simulations and are therefore beyond the scope of this initial paper on this subject.  

To conclude, the assumption of a switch-on/off mechanism for radio halos by CR transport mechanism, which operates on timescales below a Gyr, is realistic. 
The above calculated CR profiles can therefore provide us with useful guidance, at least for the cluster centers. 
\begin{figure*}
 \includegraphics[bb=25 180 550 600,width=0.49\textwidth]{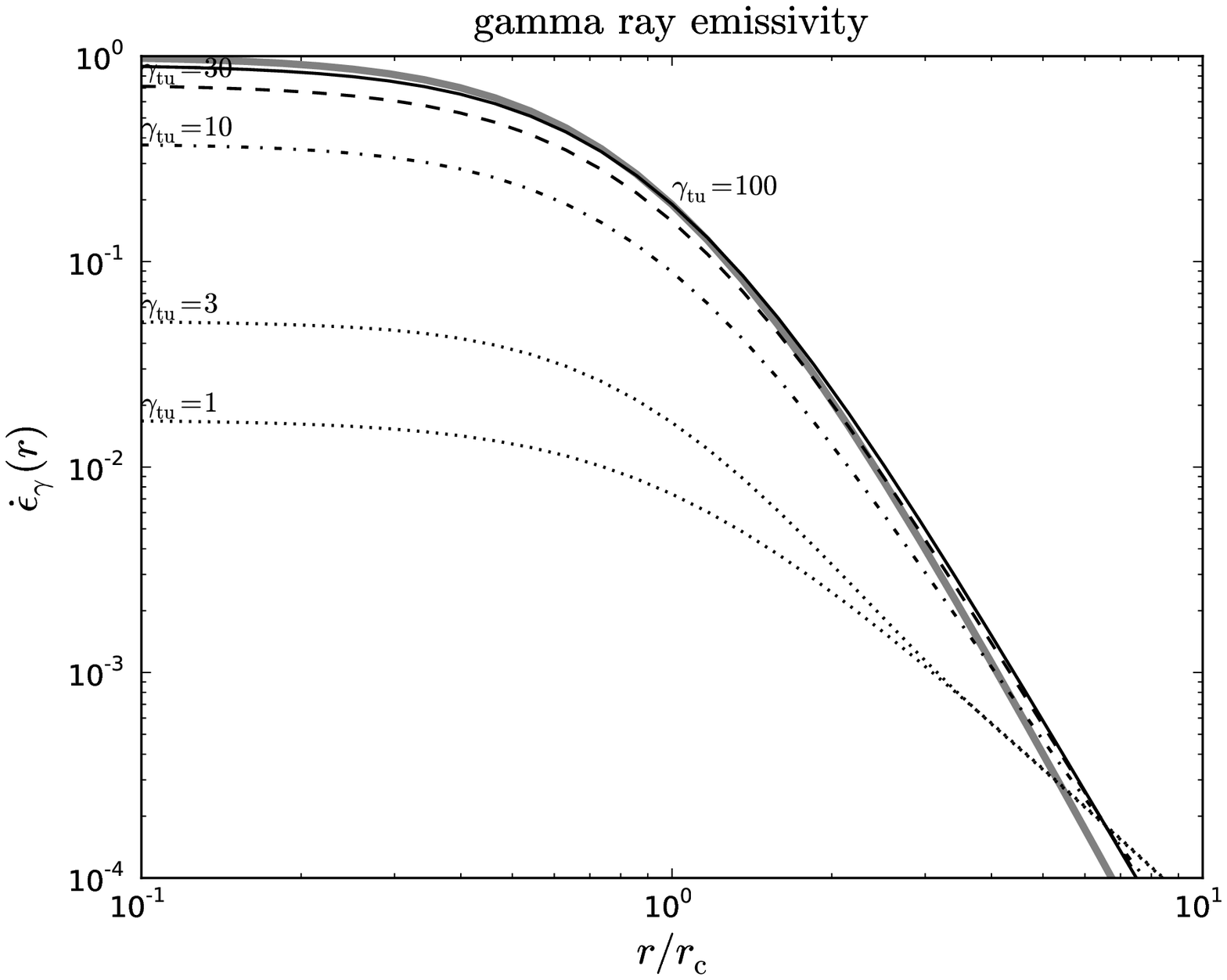}
 % density.eps: 0x0 pixel, 300dpi, 0.00x0.00 cm, bb=18 180 594 612 
\includegraphics[bb=25 180 550 600,width=0.49\textwidth]{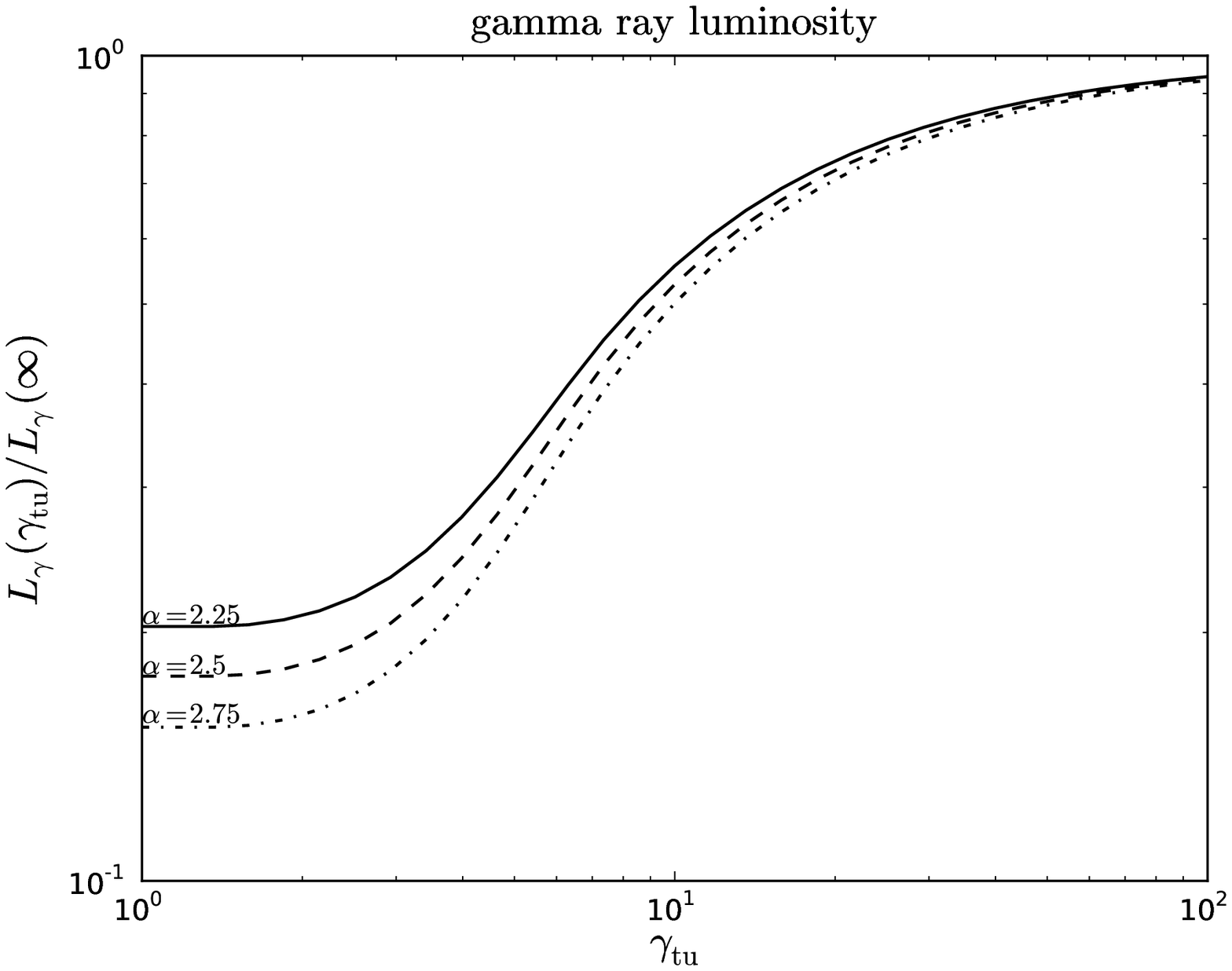}
 % density.eps: 0x0 pixel, 300dpi, 0.00x0.00 cm, bb=18 180 594 612
 \caption{\textbf{Left:} Gamma-ray emissivity profiles for the CR distributions in Fig. \ref{fig:density} and X-ray emissivity profile of the ICM in grey. Emissivities are normalised to the central emissivity of a cluster with $\gamma_\rmn{tu}=\infty$. 
\textbf{Right:} Total gamma-ray flux due to hadronic CRp interactions with the ICM nucleons as a function of $\gamma_\rmn{tu}={\tau}_\rmn{st}/\tau_\rmn{ad}$ and for $\alpha = 2.25$, $2.5$, and $2.75$ (solid, dashed, and dashed-dotted lines, respectively). 
  Normalised to $L_\gamma$ for $\gamma_\rmn{tu}=\infty$. ($\beta_\rmn{cl}=0.8$, $R_\rmn{as} =10$).}
 \label{fig:gamma}
\end{figure*}

\section{Implications for thermal and non-thermal emission}
\label{sec:implication}

In the following we illustrate the consequences of CR transport for the gamma-ray and radio halo emission of clusters, and for the heating of cluster cool cores using the two scenarios described above: confined and mobile CRs.

\subsection{Gamma-rays}
\label{sec:gamma}

The gamma-ray emissivity of a power law CRp spectrum as in Eq. \ref{eq:CRspec} is
\begin{equation}
 \lambda_\gamma = A_\gamma\, C\, \varrho_\rmn{gas},
\end{equation}
where $A_\gamma$ is provided in App. \ref{sec:gammaApp} and depends on the spectral index and the gamma-ray energy window considered, and $\varrho_\rmn{gas}$ is the gas mass density. 
The gamma-ray emissivity of a cluster with constant metallicity ($\varrho_\rmn{gas}(r) = \mu\, n_\rmn{e}(r)$, with $\mu$ the gas mass per electron) is therefore 
\begin{equation}
 L_\gamma = \int dV\, \lambda_\gamma = 4\,\pi \, A_\gamma\, \mu\, \int_0^{R_\rmn{shock}}\!\!\!\!\!\!\! dr\, r^2\, C(r)\, n_e(r).
\end{equation}
If we now assume the case of confined CRp, which expand upon vanishing turbulence by a factor $X \approx 2$, we find the gamma-ray emissivity to be reduced by the factor
\begin{equation}
 \frac{L_\gamma'}{L_\gamma}= X^{- \frac{\alpha+2}{3}} \approx 2^{-1.5} \approx 0.35.
\end{equation}
We note that the effect of lowering the target density at the location of the relaxed CRs is exactly canceled by increase in volume. Thus, only a moderate reduction of the gamma-ray flux has to be expected, which is good news for the gamma-ray detectability of galaxy clusters.

In case of mobile CR, the difference in gamma-ray luminosity between turbulent and quiet clusters is expected to be  larger, but apparently less than an order of magnitude, as can be seen from Fig. \ref{fig:gamma}. 

\subsection{Radio halos}
\label{sec:radioHaloes}

\subsubsection{Hadronic model}

Hadronic CRp collisions produce also charged pions, which decay into relativistic $e^\pm$. These cool due to   inverse Compton scattering on CMB photons and synchrotron losses and establish a steady state spectrum which is steeper than that of the parent protons by one, $\alpha_\e= \alpha+1$. The radio synchrotron emission of these electrons has a spectrum with index $\alpha_\nu = (\alpha_e -1)/2 =\alpha/2 \approx 1.15 \ldots 1.25$.
The radio luminosity at frequency $\nu$ per unit frequency interval is
\begin{eqnarray}\label{eq:Lnu}
  L_\nu &=& 
  A_\nu \int\dd V\,C_\p \, \varrho_\rmn{gas}\, \frac{\eps_B}{\eps_B+\eps_\rmn{ph}}
  \left(\frac{\eps_B}{\eps_{B_\rmn{c}}}\right)^{\frac{\alpha-2}{4}},
\end{eqnarray}
where the abbreviations $A_\nu$  and $\eps_{B_\rmn{c}}$ are defined in App. \ref{sec:radioApp}. 
These quantities are unchanged by CRp transport processes, and can be regarded to be constant, in our context.
$\eps_\rmn{ph}$ is the energy density of the photon field on which the CRe scatter. 
If this is dominated by the CMB, it corresponds to an equivalent field strength of $B_\rmn{cmb} = 3.27 \, (1+z)^2 \, \mu$G, where $z$ is the redshift. At locations where the ICM fields are stronger than this, synchrotron cooling dominates, and the radio emissivity depends only weakly on the magnetic fields in which the CRp reside. Where $B$ is below $B_\rmn{cmb}$, IC scattering is the dominant loss channel, and the radio flux is very sensitive to the field strength. 

When the cluster turbulence decays, the CRp occupy mostly the weaker magnetised regions of their enclosing field structures, in the confined CR scenario. Furthermore, magnetic fields tend to become weaker, since the magnetic small scale dynamo maintaining the fields is less powerful. We model these effects by 
\begin{equation}
 \eps_B \rightarrow \eps_B' = \eps_B \, X^{-\frac{4}{3}}\, X_\rmn{tu}^{-1}, \mbox{~with~}X_\rmn{tu} = \frac{\eps_\rmn{tu}'}{\eps_\rmn{tu}} = \frac{{\alpha'}^2_\rmn{tu}}{\alpha^2_\rmn{tu}},
\end{equation}
 the ratio of the turbulence after and during the merger. 
In case the magnetic fields are relatively weak ($B\ll B_\rmn{cmb}$), the cluster radio luminosity changes due to the CR profile relaxation according to
\begin{equation}
 \frac{L_\nu'}{L_\nu} \approx X^{-\frac{2}{3}\,(\alpha+2)} \, X_\rmn{tu}^{-\frac{\alpha+2}{4}} \approx X^{-3}\, X_\rmn{tu}^{-1.125} \approx 0.06
\end{equation}
In the approximations, we used $\alpha = 2.5$,  $X\approx 2$, and $X_\rmn{tu} = 2$. The latter is probably an underestimate of the turbulence ratio, see \citet{2009MNRAS.393.1073B}. 
If the cluster radio emission is dominated by regions with strong magnetic fields  ($B\gg B_\rmn{cmb}$), we find
\begin{equation}
 \frac{L_\nu'}{L_\nu} = X^{- \frac{4}{3}} \, X_\rmn{tu}^{-\frac{\alpha-2}{4}} \approx X^{-1.333}\, X_\rmn{tu}^{-0.125} \approx 0.36.
\end{equation}
Thus, in the confined CR scenario a radio flux decrement of at least a factor three, but more plausibly by a larger margin, can be expected.

\begin{figure*}
 \includegraphics[bb=25 180 550 600,width=0.49\textwidth]{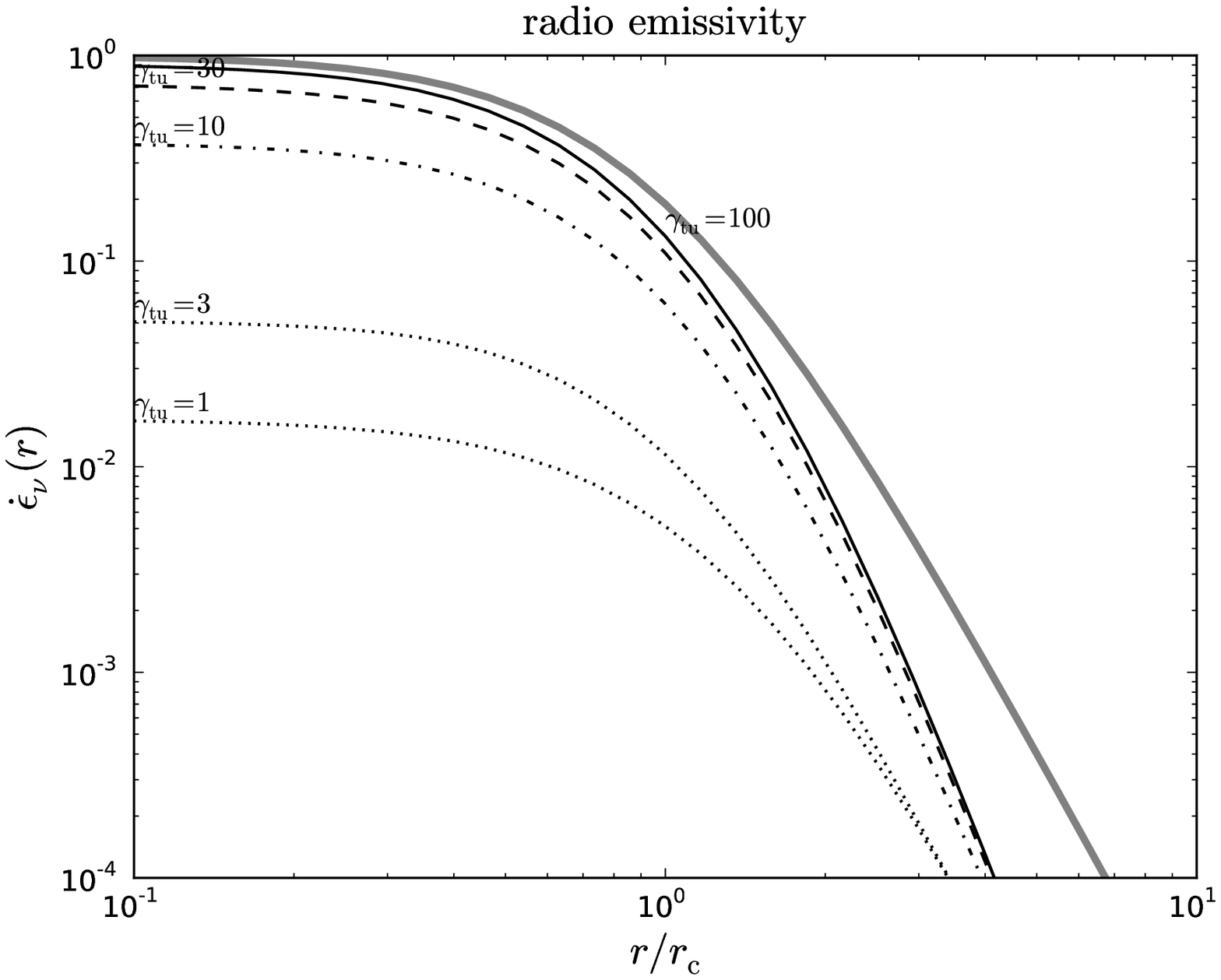}
 \includegraphics[bb=25 180 550 600,width=0.49\textwidth]{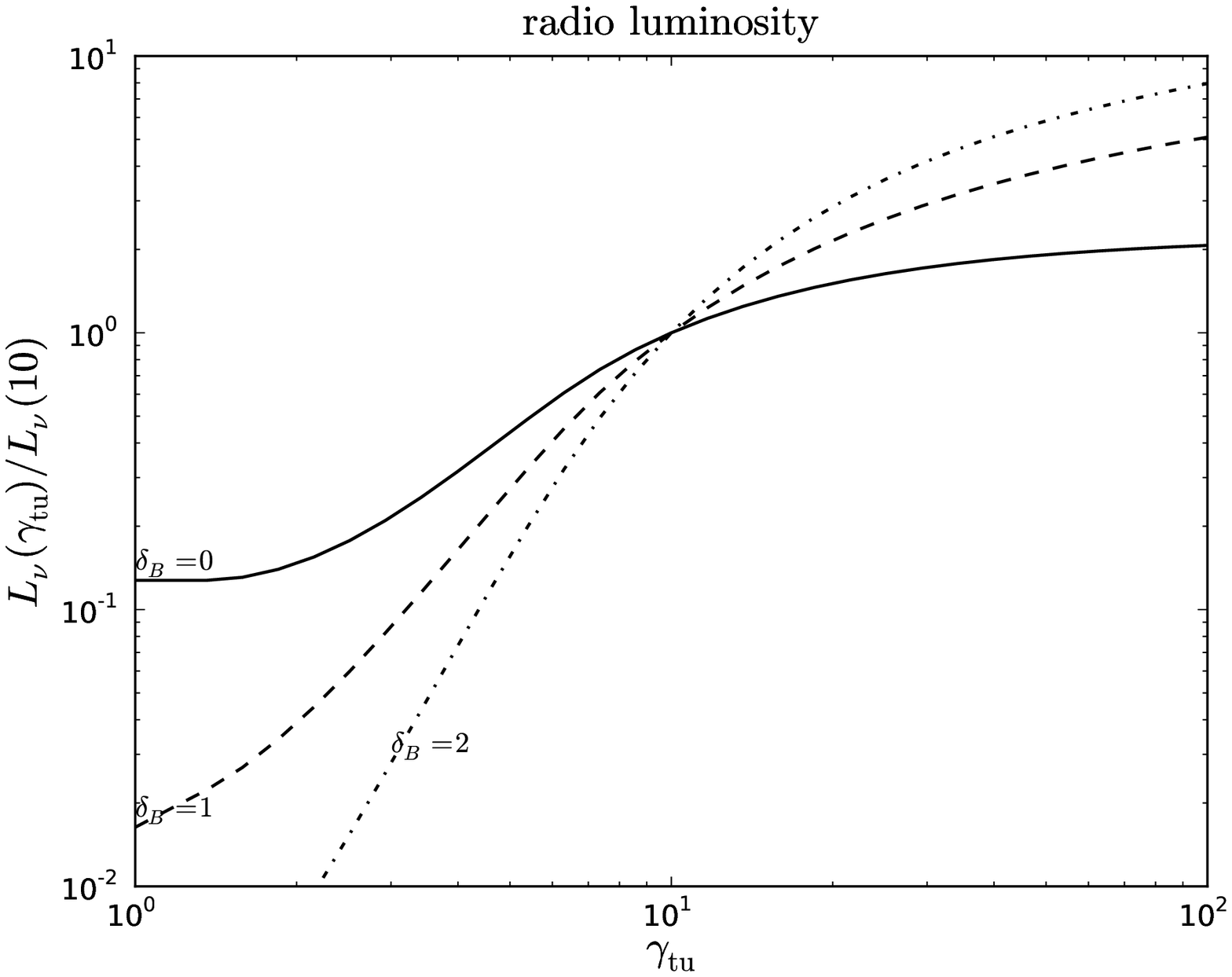}
 % density.eps: 0x0 pixel, 300dpi, 0.00x0.00 cm, bb=18 180 594 612
 \caption{\textbf{Left:} Radio emissivity profiles for the cluster shown in Fig. \ref{fig:density} assuming the same magnetic field profiles with $B_0 = 6\,\mu\rmn{G}$ and $\delta_B = 0$. Emissivities are normalised to the central radio emissivity of a cluster with $\gamma_\rmn{tu}=\infty$. The X-ray profile is shown in grey. \textbf{Right:} Total radio flux due to hadronic CRp interactions with the ICM nucleons as a function of $\gamma_\rmn{tu}={\tau}_\rmn{st}/\tau_\rmn{ad}$ and for different dependencies of the magnetic field strength on the turbulence level as parametrised by $\delta_B = 0$, $1$, and $2$ (solid, dashed, and dashed-dotted lines, respectively).  
  Normalised to $L_\nu$  for $\gamma_\rmn{tu}=10$. ($\beta_\rmn{cl}=0.8$, $R_\rmn{as} =10$, $B_0 = 6\,\mu\rmn{G}$, $\alpha=2.5$, $z=0$)}
 \label{fig:radio}
\end{figure*}

In case of a mobile CRp population, the variation of the radio flux as a function of turbulence strength is much larger. It depends on the magnetic field strength, and especially on how much the magnetic fields change in response to a change of the turbulence strength. Magnetic field generated by a saturated state of a small scale dynamo should have $\eps_B \propto \eps_\rmn{tu} \propto \alpha_\rmn{tu}^2$. However, since the magnetic field strength probably requires some time to reach the saturation state, we model also weaker dependencies by adopting
\begin{equation}
 \eps_B(r) = \frac{B_0^2}{8\,\pi}\, \frac{n(r)}{n_0}\, \left( \frac{\gamma_\rmn{tu}}{10}\right)^{\delta_B},
\end{equation}
with $B_0 = 6\,\mu$G and using $\delta_B = 0$, $1$, as well as $2$ for no, a moderate and a strong dependence of the magnetic field strength on the turbulence level, respectively. 
Since we expect the streaming velocity $\vel_\rmn{st}$ to depend strongly on the turbulence level in an inverse fashion,  $\gamma_\rmn{tu}=\vel_\rmn{tu}/\vel_\rmn{st}$ might change much stronger than $\vel_\rmn{tu}$, and therefore we expect $\delta_B < 2$.
The resulting radio luminosities are shown in Fig. \ref{fig:radio}.

It is obvious from this discussion that a rapid drop in radio luminosity after the turbulent merger phase by one order of magnitude or more, can easily be achieved; moreover, it is actually expected. 

\subsubsection{Re-acceleration model}

The presence, luminosity, and spectrum of a radio halo in the re-acceleration model depends on two things: the availability of a low energy CRe pool and the strength of the wavefield providing the CRe acceleration. To calculate the latter is a complex task, well beyond the scope of this paper. For magnetosonic waves it is calculated in \cite{2007MNRAS.378..245B, 2010arXiv1008.0184B}, where it is shown, that if a significant faction of about 20\% of the cluster pressure is in form of such waves (and their spectrum is Kraichnan-like and reaches down to small scales), electrons are re-accelerated up to 10~GeV, the radio observable part of the CRe spectrum in clusters. It should just be noted, that the generation of magnetosonic waves with the Lighthill mechanism  scales with roughly the second power of the turbulence energy density \citep{1952RSPSA.211..564L, 1954RSPSA.222....1L} and therefore can be expected to vary largely between merging and relaxed clusters. 

Since the radio emitting CRe loose their energy within 0.1 Gyr, an arbitrary strong decline of the radio power is possible in this model, if no CRp population is present replenishing CRe via the hadronic mechanism. 
In an quiescent phase of the cluster, any low energy CRe population in the cluster center is probably completely diminished by the severe Coulomb losses (see Fig. \ref{fig:ecooltime}), and has to be re-established by advection from the outer regions with less severe Coulomb losses during the next turbulent merger. 
Thus, large variations of the radio luminosity in the re-acceleration model can be expected, as shown by \cite{2009A&A...507..661B}.

\subsection{Cool core heating}

\begin{figure}
 \includegraphics[bb=25 180 550 600,width=0.49\textwidth]{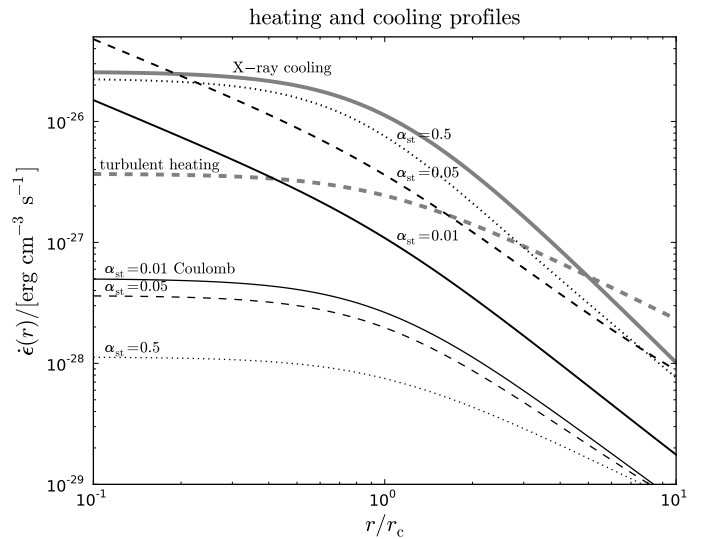}
 % density.eps: 0x0 pixel, 300dpi, 0.00x0.00 cm, bb=18 180 594 612
 \caption{CR heating profiles due to CR streaming (upper black curves) and Coulomb losses (lower black curves)  for the CR distributions resulting from different streaming velocities as indicated at curves, the turbulent heating profile (grey dashed line, calculated according to \citet{2007A&A...473...41E}), and the X-ray cooling profile (grey solid line). 
The adopted cool core parameters are $n_0 =0.05\; \rmn{cm^{-3}}$, $kT=3\,$keV, $r_\rmn{c}= 10\,$kpc, $\vel_\rmn{tu}= 350 \,\rmn{km\,s^{-1}}$, $\beta_\varrho = 1$, $\beta_\rmn{cc}=0.4$, $R_\rmn{cc} =10\,r_\rmn{c}$, $\alpha=2.5$, and a CR to thermal energy ratio of $\eps_\rmn{cr}/\eps_\rmn{th} = 0.15$ at the outer radius of the cool core, $R_\rmn{cc} $, and less in the center (down to $\eps_\rmn{cr}/\eps_\rmn{th}\approx 0.01$ for large $\alpha_\rmn{st}$).
Since we assume the gas motions to be mostly convective, we set $\chi_\rmn{tu} = 10$ to model an efficient radial transport. 
This implies $\alpha_\rmn{tu} = 0.36$ and $\kappa_\rmn{tu}= 3.6\,10^{30}\,\rmn{cm^2\,s^{-1}}$.  
We further have $\gamma_\rmn{tu} = 2.4,$ $24,$ and $120$ for $\alpha_\rmn{st}= 0.5$, 0.05, and 0.01, respectively. 
The total X-ray cooling luminosity of the cool core region is about $4.4\,10^{43}\,\rmn{erg\,s^{-1}}$. 
The heating power of the turbulence is about the same, but mostly going into outskirts of the cool core. 
The total CR heating power (streaming plus Coulomb) within the $R_\rmn{cc}= 10\, r_\rmn{c}$ is 2.9, 2.4, and 0.8~$10^{43}\,\rmn{erg\,s^{-1}}$ for $\alpha_\rmn{st} = $0.5, 0.05, and 0.01, respectively.}
 \label{fig:heatcool}
\end{figure}

\begin{figure}[t]
 \includegraphics[bb=25 180 550 600,width=0.49\textwidth]{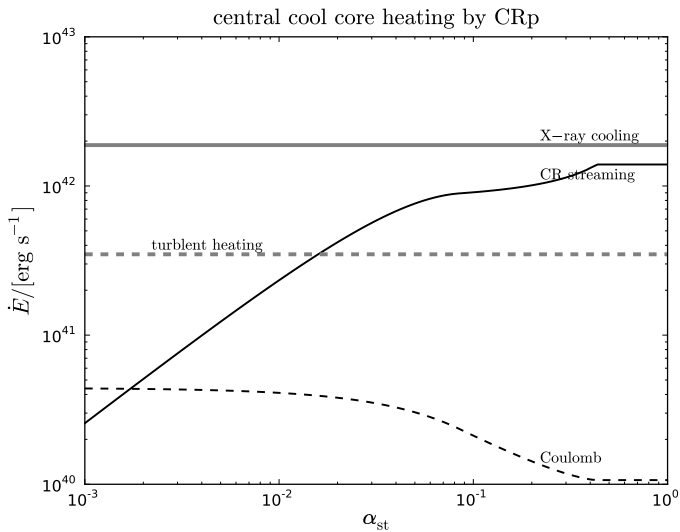}
 % totalheat.eps: 1179666x1179666 pixel, 300dpi, 9987.84x9987.84 cm, bb=18 180 594 612
 \caption{Central CR heating within $r_\rmn{c}$ of the cool core in Fig. \ref{fig:heatcool} as a function of the CRp streaming velocity. The CR streaming heating power decreases with decreasing streaming velocity. For $\alpha_\rmn{st}\sim 0.1$ this trend is weakened by the build up of a centrally enhanced CRp population. For $\alpha_\rmn{st}\sim 1$ the streaming velocity is so large, that the advective work on the CRp population is instantaneously dissipated, and the heat input is therefore independent of $\alpha_\rmn{st}$ in this regime. The build up of the central CR population with decreasing $\alpha_\rmn{st}$ can be seen in the Coulomb heating rate.}
 \label{fig:cctotalheat}
\end{figure}

The energy lost by CR during streaming heats the ICM, since the plasma waves excited by the CR dissipate. The origin of this energy is, in absence of CR sources pumping fresh relativistic particles into the ICM, the kinetic energy of ICM gas flows. Thus, the heating due to CR streaming is not providing an additional heat source to the ICM, since the kinetic energy of the gas will be dissipated via a turbulent cascade anyway otherwise, but CR streaming provides a different channel for this energy dissipation, with a different spatial footprint. 

As we show in the following, the heating profile of CR streaming is centrally concentrated, and therefore may play a role in stabilizing cool core regions of galaxy clusters. In cool cores, the coldest gas is at the center and is therefore also the densest, leading to the shortest cooling time. In order that no cooling catastrophe occurs, as observations indicate, this central gas has to be heated preferentially.

In order to get an idea on the heating profile due to CR streaming, we adopt the simple picture for the CR cluster profile of Sect. \ref{sec:expectedCRprofile} for a model of an individual cool core. Thus we describe the cool core with a beta-profile with small $r_\rmn{c} \sim 10$ kpc. However, instead of assuming a fixed number of CRs within the cool core region, we assume that the non-cool core region provides an environmental CR floor with fixed CR density $\varrho_\infty$, to be dragged into the cool core via the gas motion therein. We still model this gas motion again as isotropic turbulence, but keep in mind, that in reality it may have more a convective structure, since it is probably excited by rising radio bubbles from the central galaxies. These bubbles are one way -- if not the dominant way --  of the central galaxy to provide energetic feedback to the cool core preventing catastrophic cooling. CRs escaping from such bubbles will add to the heating, as discussed in the references given in footnote \ref{note:coolcoreheat}. We ignore here this contribution, in order to see the contribution to cool core heating from the interplay of CR advection and streaming alone.

The energy deposition of a radially streaming CR with momentum $p\,m\,c$ is
\begin{eqnarray}
 \dot{E}_\mathrm{cr}(p) &=& \frac{\partial E_\mathrm{cr}(p)}{\partial p} \dot{p} = - \frac{p^2 \,m\,c^2}{3\,\sqrt{1+p^2}}\, \vec{\nabla} \cdot \vvst \nonumber\\
&=&  - \frac{2\, p^2 \,m\,c^2}{3\,\sqrt{1+p^2}}\, \left(\frac{\vst}{r} + \frac{1}{2}\, \frac{\partial \vst}{\partial r} \right)
\end{eqnarray}

If we integrate this over the spectrum given in Eq. \ref{eq:CRspec} (and thereby restricting our discussion to the energetically more important CRp), while introducing a low momentum cutoff $p_0$, to be on the safe side, we find the heating power to be
\begin{equation}
 \dot{E}_\mathrm{st} = \frac{C(r)\,m\,c^2}{3}\, \mathcal{B}_{q} \left( 
\frac{\alpha -2}{2}, \frac{3-\alpha}{2}\right) \, \left(\frac{\vst}{r} + \frac{1}{2}\, \frac{\partial \vst}{\partial r} \right) ,
\end{equation}
with $\mathcal{B}_{q}(a,b)$ denoting the incomplete Beta-function, and $q=(1+p_0^2(r))^{-1}$. Within $r_- < r< r_+$ we can drop the $\partial \vel_\mathrm{st}/\partial r $ term, since there $\vel_\rmn{st} = \rmn{const}$. Outside of this range, the steady state solution requires 
\begin{displaymath}
\vel_\rmn{st} =- \vel_\rmn{ad} = \kappa_\rmn{tu}\, \frac{\partial}{\partial_r}\, \ln\left(\frac{\varrho}{\eta}\right) =  -\kappa_\rmn{tu}\, \frac{\partial\, \eta}{\eta\, \partial_r} = \frac{3\,\beta_\rmn{cl}\, \kappa_\rmn{tu}}{\gamma\,r_\rmn{c}}\, \frac{r/r_\rmn{c}}{1+r^2/r^2_\rmn{c}},
\end{displaymath}
since we assumed a constant (time-averaged) CRp density profile there.

The low momentum cutoff%
\footnote{This cutoff can can, however,  be well set to $p_0=0$ in a simplified treatment as long as $\alpha$ is sufficiently smaller than $3$, which should be the case for the expected $\alpha=2.5$. In this case
\begin{displaymath}
 \dot{\eps}_\mathrm{st} = \frac{C(r)\,\vst\,m\,c^2\,\Gamma^2\left(\frac{1}{4}\right)}{3\,\sqrt{\pi}\,r}\approx 2.5 \, \frac{C(r)\,\vst\,m\,c^2}{r}
\end{displaymath}
for $r_-<r<r_+$.
The numerical factor in the last expression changes to $1.2$ if a cut off at $p_0=1$ is used, which is typical for CRp under the influence of Coulomb losses depopulating the sub-relativistic regime. } 
depends weakly on the position via $p_0(r) = p_0 \, (n(r)/n_0)^{1/3}$. We use $p_0(R_\rmn{cc})=1$, which roughly mimics the Coulomb cooling cutoff the CRp spectrum develops \citep{2007A&A...473...41E}.

As can be seen in Fig. \ref{fig:heatcool} CRp heating is strongly enhanced centrally. 
This is not only due to the central peak of the CRp density, but mostly due to the $1/r$ factor caused by the streaming for $r_-<r<r_+$. But even for completely flat CRp profiles (see curve $\alpha_\rmn{st}=0.5$), where this $1/r$-factor does not apply, there is a centrally enhanced heating. This results from the $PdV$-work done by the gas motion while trying to drag in  the CRs, which is instantaneously dissipated via streaming. The CRp heating  decreases with decreasing streaming speed, since without streaming no energy can be dissipated through the excitation of plasma waves.  

Thus, the CRs collect some fraction of the kinetic energy distributed through the cool core, and release it preferentially in the cool core center, whereas turbulent cascades just heat each mass element with the same rate (in our simplified turbulent cool core model):
\begin{equation}
  \dot{\eps}_\mathrm{tu}(r) = \frac{{E}_\mathrm{tu}(r)}{\tau_\mathrm{tu}} = \frac{\vel_\rmn{tu}^3}{\pi\,L_\rmn{tu}}\,\varrho_\mathrm{gas}(r)
\end{equation}

 The parameters of our cool core example in Fig. \ref{fig:heatcool} were chosen to indicate that CRp heating is a potentially interesting mechanism. We could have picked parameters for which this heating would be dominant or insignificant in the center (whereas the latter is more easily achieved). It is not clear to us at the moment if the self-regulation processes in cool cores drive the system to a state where CRp streaming heating is significant or not. This depends to some degree on the value of $\alpha_\rmn{st}$, which we do not know for this environment, as well as on the amount of CRp in the surrounding of the cool core and other cool core parameters. The dependence of the central CRp heating rate on $\alpha_\rmn{st}$ is shown in Fig. \ref{fig:cctotalheat} for the same cool core parameters. For a sufficiently large streaming speed ($\alpha_\rmn{st} > 0.05$) the heating rate due to streaming is actually relatively insensitive to this speed and dominates the central heating in this particular cool core.

In order to see under which circumstances CRp streaming can be an significant heating mechanism, we have to compare the dependencies of the cooling and heating rates with cool core parameters. To simplify the discussion, we ignore line cooling (as we did in Figs. \ref{fig:heatcool} and \ref{fig:cctotalheat}) and assume maximally efficient streaming and thus a spatially flat CRp profile within the cool core region. With these simplifications, the profiles of  X-ray cooling $\dot{\eps}_X(r)$, of turbulent heating $\dot{\eps}_\rmn{tu}(r)$, and the heating by CRp streaming $\dot{\eps}_\rmn{tu}(r)$, are given by
\begin{eqnarray}
 \dot{\eps}_X(r) &=& -\Lambda_0 \, (kT)^{\frac{1}{2}}\, n_\rmn{0}^2\,\left( 1+ r^2/r_\rmn{c}^2\right)^{-3\,\beta_\rmn{cc}}\nonumber\\
\dot{\eps}_\rmn{tu}(r) &=& \frac{\mu\,n_\rmn{e,0}\,\vel_\rmn{tu}^3}{\pi \, \chi_\rmn{tu}\,r_\rmn{c}}\, \left( 1+ r^2/r_\rmn{c}^2\right)^{-\frac{3}{2}\,\beta_\rmn{cc}}\\
\dot{\eps}_\rmn{st}(r) &=& \frac{\beta_\rmn{cl}\, \vel_\rmn{tu}\,\chi_\rmn{tu}}{\gamma\, r_\rmn{c}}\, P_\rmn{cr,0}\,
\frac{3+ r^2/r_\rmn{c}^2}{( 1+ r^2/r_\rmn{c}^2)^{2}}.\nonumber
\end{eqnarray}
Here we used the formula for the pressure of a CRp spectrum with a power-law momentum distribution as given in \citet{2007A&A...473...41E}, wrote $\mu \approx m_\rmn{p}$ for the mean molecular weight per electron, and introduced the X-ray cooling constant $\Lambda_0 = 5.96\,10^{-24}\,\rmn{erg\, s^{-1}\,cm^3\,(keV)^{-1/2}}$ for a metallicity of 0.3 solar.

Heating by CRp streaming can therefore become comparable to the cooling in the center if the CRp pressure is relatively high, turbulence is strong, the cool core radius is small, and the electron density is low. The ratio of CRp streaming heating and X-ray cooling scales as
\begin{equation}
 \frac{\dot{\eps}_\rmn{st}(0)}{|\dot{\eps}_X(0)|} \propto \frac{P_\rmn{cr,0}}{P_\rmn{th,0}}\, \frac{\beta_\rmn{cc}\,\alpha_\rmn{tu}\chi_\rmn{tu}\, kT}{n_\rmn{e}\,r_\rmn{c}},
\end{equation}
whereas that with turbulent heating scales as
\begin{equation}
  \frac{\dot{\eps}_\rmn{st}(0)}{\dot{\eps}_\rmn{tu}(0)} \propto \frac{P_\rmn{cr,0}}{P_\rmn{th,0}}\, \frac{\beta_\rmn{cc}\,\chi_\rmn{tu}^2}{\alpha_\rmn{tu}^2}.
\end{equation}

From this we expect that heating by CRp streaming is most likely to be of importance in weak cool cores, where the central electron density is not that extreme, the cool core radius is small, and the temperature is still relatively high. A strong convective structure of the turbulence with $\chi_\rmn{tu}>1$ is also very beneficial for heating by CRp streaming, especially to boost it in comparison to turbulent heating. Thus, we speculate that CR streaming might moderate the initial growth of cool cores, whereas a massive cool core is more likely stabilised by the dissipation of stronger turbulence and other processes not discussed here (dissipation of weak shock waves, radiative heating, ...).
The cool core parameters of Figs. \ref{fig:heatcool} and \ref{fig:cctotalheat} were actually chosen to represent such a case of a weak cool core with strongly convective turbulence.

Note, that if CRp diffusion would also be a significant transport mechanism in cool cores, we expect some level of Fermi~I CRp acceleration to take place, as discussed in Sect. \ref{sec:acc}. This might even allow that a sufficiently strong CRp population builds up automatically within the cool core to counterbalance central cooling, even if the seed CRp population was not very energetic. If such an energetic population could be established this way depends also on the escape rate of CRp from the cool core, a quantity we do not know with our limited knowledge on the magnetic topology in such environments. One might speculate if the presence of radio-mini halos in some cool cores has something to do with this possible CR acceleration mechanism.

\section{Discussion}
\label{sec:discuss}

We argue that streaming is an important CR transport mechanism in galaxy clusters. It tends, as CR diffusion, to establish a spatially flat CR profile, and therefore to drive CRs out of the cluster core. This can explain why radio halos are not found in every cluster, although simulations indicate that a sufficient number of CR protons should have been accumulated in clusters for this.

CR advection, on the other hand, tends to produce centrally enhanced CR profiles, as any turbulent mixing process in a stratified atmosphere tries to establish a constant abundance profile, so that the density of the advected quantity follows the density of the fluid.

Thus, CR advection and streaming are counteracting transport mechanisms. 
Whenever the former dominates, centrally enhanced profiles are established, and whenever streaming is more important, a flat profile results. 
The crucial quantity is the advective-to-streaming-velocity ratio, $\gamma_\rmn{tu}=\vel_\rmn{tu}/\vel_\rmn{st}$.

The advective velocities in galaxy clusters are comparable to the sound speed during cluster merger, and less when the cluster relaxes after the merger. 
The streaming velocity is poorly known. 
On a microscopic scale, it might be of the order of the sound speed, or even much larger, if the plasma wave turbulence level is low. 
Macroscopically it might be reduced by a large factor from its microscopic value, due to magnetic trapping of CR in flux tubes and the slow cross field diffusion required to escape. Also this topological reduction factor for the streaming speed should depend on the level of turbulence that is present in clusters.

Thus, there are three factors simultaneously increasing $\gamma_\rmn{tu}=\vel_\rmn{tu}/\vel_\rmn{st}$ when the cluster turbulence increases: the turbulent velocities increase, the microscopic streaming speed decreases due to larger level of plasma waves, and the macroscopic streaming speed is further decreased due to a more complex magnetic topology. Taken together, the combination of these effects should produce a significant variation of $\gamma_\rmn{tu}$ between merging and relaxed clusters.

As a result of this, merging clusters should have a much more centrally concentrated CR population than relaxed ones. This leads naturally to a bimodality of their gamma-ray and radio synchrotron emissivities due to hadronic interactions of CR protons. Also in the re-acceleration model of cluster radio halos these transport processes should be essential, since the re-accelerated CR electron populations in the dense cluster centers is probably too vulnerable to Coulomb losses, to survive periods without significant re-acceleration. Transport of the longer living electrons at the cluster outskirts into the cluster center during cluster merger would circumvent this problem.

Although we did not work his out in detail, it should be noted that we expect an energy dependence of the macroscopic CR streaming speed, which then should lead to a spatial differentiation of the spectral index of the CRp population and any secondary radio halo emission. Such spectral index variation in the radio halo should become especially strong during phases of outstreaming CRp, i.e. when a radio halo dies due to the decay of the cluster turbulence.

We have also shown that CR streaming in cluster cool cores can help to dissipate the turbulent energy preferentially in the cool core center, and thereby potentially help to stabilise cool cores against a cooling instability. The CR population in cool cores required for this could be either dragged into the cool core by turbulent transport or be self-maintained by ongoing Fermi~I acceleration if CR diffusion is also an important CR propagation mode.

To conclude, we have shown that CR transport mechanisms seem to be essential to understand the non-thermal content and radiative signatures of clusters. Additionally, they might play a role in shaping the clusters' thermal structure, especially in cool cores. Further investigations of CR transport in galaxy clusters using detailed three-dimensional numerical simulation of the involved processes are therefore necessary as well as an improved understanding of the dependence of the macroscopic CR streaming velocity on cluster weather conditions.

\begin{acknowledgements}
We thank Henrik Junklewitz and an anonymous referee for comments on the manuscript and Anvar Shukurov for good company during several of our discussions. All numerical calculations were done with the open source package Sage Version 4.3.3 \citep{sage}. This research was performed in the framework of the DFG Forschergruppe 1254 ``Magnetisation of Interstellar and Intergalactic Media: The Prospects of Low-Frequency Radio Observations''.
\end{acknowledgements}

\appendix

\section{Gamma-ray emissivity}
\label{sec:gammaApp}

The gamma-ray emissivity in photon number between the energies $E_1$ and $E_2$ of a power law CRp spectrum as in Eq. \ref{eq:CRspec} is according to \cite{2008MNRAS.385.1211P} (which we follow here closely)
\begin{eqnarray}
\label{eq:lambda_gamma}
\lambda_\gamma &=& \lambda_\gamma(E_1, E_2)  = \int_{E_1}^{E_2} \dd E_\gamma\,
s_\gamma(E_\gamma) \\
\label{eq:lambda_gamma2}
&=& \frac{4\, C}{3\, \alpha\delta_\gamma}
\frac{m_{\pi^0} c\, \sigma_{\rmn{pp}}n_\rmn{N}}{m_\p}
\left( \frac{m_\p}{2 m_{\pi^0}}\right)^{\alpha}\,
\left[\mathcal{B}_x\left(\frac{\alpha + 1}{2\,\delta_\gamma},
    \frac{\alpha - 1}{2\,\delta_\gamma}\right)\right]_{x_1}^{x_2},\\
x_i &=& \left[1+\left(\frac{m_{\pi^0}c^2}{2\,E_i}
      \right)^{2\,\delta_\gamma}\right]^{-1} \mbox{~for~~}
      i \in \{1,2\},
\end{eqnarray}
where $n_\rmn{N} = n_\rmn{H} + 4 n_\rmn{He} = \rho_\rmn{gas} /
m_\p$ is the target density of nucleons in a fluid of primordial element
composition,
\begin{equation}
 \label{eq:sigmapp}
\sigma_\rmn{pp} \simeq 32 \cdot
\left(0.96 + \rmn{e}^{4.4 \,-\, 2.4\,\alpha_\p}\right)\mbox{ mbarn}
\end{equation}
the spectrally weighted hadronic cross section, $m_{\pi^0} c^2 \simeq 135 \,\mbox{MeV}$ the rest mass of a neutral
pion, and the shape parameter
\begin{equation}
\label{eq:delta}
\delta_\gamma \simeq 0.14 \,\alpha_\gamma^{-1.6} + 0.44.
\end{equation}
There is a detailed discussion in \citet{2004A&A...413...17P} how the
$\gamma$-ray spectral index $\alpha_\gamma$ relates to the spectral index of
the parent CRp population $\alpha$. Here, we assume $\alpha_\gamma = \alpha$. Thus we find
\begin{equation}
 A_\gamma =  \frac{4}{3}
\frac{m_{\pi^0} c\, \sigma_{\rmn{pp}}}{\alpha\, \delta_\gamma}
\left( \frac{m_\p}{2 m_{\pi^0}}\right)^{\alpha}\,
\left[\mathcal{B}_x\left(\frac{\alpha + 1}{2\,\delta_\gamma},
    \frac{\alpha - 1}{2\,\delta_\gamma}\right)\right]_{x_1}^{x_2}.
\end{equation}

\section{Hadronic radio-synchrotron emissivity}
\label{sec:radioApp}

The coefficient in Eq. \ref{eq:Lnu} characterizing the radio-synchrotron emissivity of  electrons injected from the decay of charged pions produced in hadronic  CRp-gas collisions is according to \cite{2008MNRAS.385.1211P} (which we follow here closely, but with slightly modified normalisation)
\begin{equation}
  A_\nu =  4\pi\, A_{E_\rmn{synch}}
  \frac{16^{2-\alpha_\e} \sigma_\rmn{pp}\, m_\e\,c^2}
   {(\alpha_\e - 2)\,\sigma_\rmn{T}\,\eps_{B_\rmn{c}} \,m_\p}
   \left(\frac{m_\p}{m_\e}\right)^{\alpha_\e-2}
   \left(\frac{m_\e c^2}{\rmn{GeV}}\right)^{\alpha_\e-1},
\end{equation}
with the dimensions $[A_\nu] =
\mbox{erg cm}^3 \mbox{ s}^{-1} \mbox{ Hz}^{-1}$ and the volume integral extends
over the entire cluster. 

The frequency $\nu$ dependent characteristic field strength $B_\rmn{c}$, and the scaling factor of the emitted energy per time $\nu^{-1}$ and frequency $A_{E_\rmn{synch}}$ are given by
\begin{eqnarray}
\label{eq:Bc}
B_\rmn{c} &=& \sqrt{8\pi\, \eps_{B_\rmn{c}}} 
   = \frac{2\pi\, m_\e^3\,c^5\, \nu}{3\,e \mbox{ GeV}^2}
   \simeq 31 \left(\frac{\nu}{\mbox{GHz}} \right)~\mu\rmn{G},
\\
A_{E_\rmn{synch}} &=& \frac{\sqrt{3\pi}}{32 \pi}
   \frac{B_\rmn{c}\, e^3}{m_\e c^2}
   \frac{\alpha_\e + \frac{7}{3}}{\alpha_\e + 1}
   \frac{\Gamma\left( \frac{3\alpha_\e-1}{12}\right)
         \Gamma\left( \frac{3\alpha_\e+7}{12}\right)
         \Gamma\left( \frac{\alpha_\e+5}{4}\right)}
        {\Gamma\left( \frac{\alpha_\e+7}{4}\right)}.
\end{eqnarray}
Here $\Gamma(a)$ denotes the $\Gamma$-function,
$\alpha_\nu = (\alpha_\e-1)/2 = \alpha/2$, and $B_\rmn{c}$ denotes a (frequency dependent)
characteristic magnetic field strength which implies a characteristic magnetic
energy density $\eps_{B_\rmn{c}}=B_\rmn{c}^2/(8\,\pi)$.

\bibliographystyle{aa}
\bibliography{cr}

\end{document}